\documentclass[aps,prl,showpacs,twocolumn,amsmath,amsfonts,amssymb,superscriptaddress,tightenlines]{revtex4-1}

\usepackage[colorlinks,citecolor=blue,linkcolor=cyan]{hyperref}
\usepackage{grffile}
\usepackage[caption=false]{subfig}
\usepackage{graphicx}
\usepackage{mathtools}
\usepackage{txfonts}

\renewcommand{\paragraph}[1]{{\par\it #1---}\ignorespaces}
\newcommand{\dg}{^{\dagger}}
\newcommand{\ndg}{^{\phantom{}}}
\renewcommand{\vec}[1]{{\boldsymbol{#1}}}
\newcommand{\op}{\hat}
\newcommand{\id}{\mathbb{I}}
\newcommand{\wt}{\widetilde}

\newcommand{\Arg}{\textsf{Arg}}

\newcommand{\pr}{^{\prime}}
\newcommand{\ua}{\uparrow}
\newcommand{\da}{\downarrow}

\renewcommand{\Im}{\textsf{Im}}

\begin{document}
\date{\today}

\title{Invariance of topological indices under Hilbert space
  truncation }

\author{Zhoushen Huang} \affiliation{Institute for Materials Science,
  Los Alamos National Laboratory, Los Alamos, NM 87545, USA}
\email{zsh@lanl.gov}
\author{W. Zhu} \affiliation{T-4 and CNLS, Los Alamos National Laboratory, Los Alamos, NM 87545, USA}
\email{zhuwei@lanl.gov}
\author{Daniel P. Arovas} \affiliation{Department of Physics, University of California San Diego, La Jolla, CA 92093, USA}
\email{darovas@ucsd.edu}
\author{Jian-Xin Zhu} \affiliation{T-4 and CINT, Los Alamos National Laboratory, Los Alamos, NM 87545, USA}
\email{jxzhu@lanl.gov}
\author{Alexander V. Balatsky} \affiliation{Institute for Materials
  Science, Los Alamos National Laboratory, Los Alamos, NM 87545, USA}
\affiliation{NORDITA, Roslagstullsbacken 23, SE-106 91\ \ Stockholm,
  Sweden}
\email{avb@nordita.org}

\begin{abstract}
  We show that the topological index of a wavefunction, computed in
  the space of twisted boundary phases, is preserved under Hilbert
  space truncation, provided the truncated state remains
  normalizable. If truncation affects the boundary condition of the
  resulting state, the invariant index may acquire a different
  physical interpretation. If the index is symmetry protected, the
  truncation should preserve the protecting symmetry. We discuss
  implications of this invariance using paradigmatic integer and
  fractional Chern insulators, $Z_2$ topological insulators, and
  Spin-$1$ AKLT and Heisenberg chains, as well as its relation with
  the notion of bulk entanglement. As a possible application, we
  propose a partial quantum tomography scheme from which the
  topological index of a generic multi-component wavefunction can be
  extracted by measuring only a small subset of wavefunction
  components, equivalent to the measurement of a bulk entanglement
  topological index.
\end{abstract}
\maketitle

\paragraph{Introduction}
The investigation of topological phases and their classification
\cite{BernevigHughes, Qi11-rmp, Ryu10, Chiu16-rmp} has grown into a
major endeavor in condensed matter physics, thanks to rapid
advancements in material realization \cite{Hasan10-rmp,Ando13} and
experimental platforms for ``quantum simulation'' such as ultra cold
atomic systems \cite{Bloch08, Bloch12,Langen15}. The appeal of
topology is that related physical quantities, for example quantized
Hall conductance \cite{TKNN} and charge polarization \cite{Zak89,
  King-Smith93}, can be formulated as discrete topological indices,
which are thus robust against continuous deformations of the
system. 

A topological index is fundamentally a property of a wavefunction.
Yet apart from free fermions and a few exactly solvable models, it is
impractical to obtain an exact wavefunction through the
diagonalization of a Hamiltonian.  One alternative is to build
candidate wavefunctions through projective construction, whereby a
parent state defined in a larger Hilbert space is linked to a
projected state in a smaller, truncated Hilbert space \cite{Suzuki80,
  Anderson87, Gros89}. Both the parent and the truncated Hilbert
spaces can play the role of the physical space.  For example, a matrix
product state is constructed by projecting a parent state, defined in
a tensor product of site Hilbert spaces, onto bond Hilbert spaces,
where truncation in bond dimension is implemented according to the
entanglement content \cite{Schollwoeck11}. In this case, the parent
space is physical, while the projected state offers a more economical
description suitable for numerical solution. In parton-type
constructions \cite{Wen02}, on the other hand, one first
fractionalizes the physical degrees of freedom into partons, with
which a mean field state can be written down in the enlarged parton
Hilbert space, then a Gutzwiller type projection is employed to pull
the state back to the physical space. In this case, the truncated
space is physical, while the enlarged space provides a more natural
platform for exotic phenomena such as fractionalization. Treated as
variational ansatz, the projected wavefunctions thus obtained can be
further optimized for better approximation of target states, yet for
the issue of topological characterization, a fundamental question
remains rarely touched: how does the truncation procedure itself
affect topology?

In this work, we investigate the connection between Hilbert space
truncation and topology on the wavefunction level. Specifically, we
address the question: what is the relation between the parent and the
projected wavefunctions in terms of their topological index?  The
topological indices we will consider are those that can be computed
via the formalism of twisted boundary phases \cite{Niu85}, such as
integer and fractional Chern numbers, quantized Berry phase, and
various symmetry protected $Z_2$ indices. We will assume that the
parent state is a gapped eigenstate $|\Psi(\vec \kappa)\rangle$ of a
many-body Hamiltonian, hence it has well-defined topological indices.
Here $\vec\kappa \equiv (\kappa_1, \kappa_2,\cdots)$ are the boundary
phases implemented as
$a^+_{\vec r + N_i \hat e_i} = a^+_{\vec r} e^{i\kappa_i}$, where
$a^+_{\vec r}$ is a fermionic/bosonic creation operator or a spin
raising operator on lattice site $\vec r$, and $N_i$ is the linear
size along direction $\hat e_i$.  The full parameter space of
$\vec \kappa$, with $\kappa_i\in [0, 2\pi)\forall i$, will be referred
to as a ``Brillouin Zone'' (BZ).  We will show that the topological
index of $|\Psi\rangle$ is fully preserved by its truncated version,
$|\wt \Psi\rangle = P|\Psi\rangle / \sqrt{\langle \Psi | P |
  \Psi\rangle}$, if both indices are computed using the \emph{same}
$\vec\kappa$ BZ, provided the $\vec \kappa$-independent projection $P$
fulfills the following conditions: (1) At no point in the $\vec\kappa$
BZ does the truncated wavefunction become a null vector, whereby
information of the parent state is fully lost.  (2) For a parent state
belonging to symmetry protected topological classes, the truncation
should also preserve the protecting symmetry in order for the
classification to remain meaningful. This is consistent with recent
works on the node structure in wavefunctions overlaps \cite{Gu16,
  Huang16}, and we discuss their relation and distinction in the
SM. Note that under certain truncation schemes, $\vec\kappa$ may no
longer correspond to physical boundary phases for the truncated
state. In such cases, truncation invariance remains true
mathematically, but acquires a different physical interpretation, and
may place the truncated state in a different topological class from
the parent state, see later discussion on the parton construction of
fractional Chern insulators.

\paragraph{Truncation invariance of Chern number and related
  topological indices}
We begin by constructively showing that the Chern number is invariant
under Hilbert space truncation. This serves as a generic proof that
any topological index obtainable from a Chern number calculation will
remain invariant under such a truncation.  Calculation of the Chern
number is at the heart of topological classification of
two-parameter-family wavefunctions.  In addition to the integer and
fractional quantum Hall effect \cite{TKNN, Kohmoto85, Niu85}, it can
also be used to classify symmetry protected topological (SPT) states
by restricting its calculation to a subset of states or a reduced
parameter space, examples include spin Chern number for
time-reversal-invariant TIs \cite{Sheng06, Fukui07, Prodan09, Yang11},
mirror Chern number \cite{Teo08} and more generally Chern numbers over
2D high symmetry manifold within a 3D single particle BZ for
crystalline TIs \cite{Alexandradinata14}. We will discuss its
implication on fractional Chern insulator states later in the text. A
step by step illustration of the proof to be discussed below can be
found in the SM using a $3$-band Hofstadter model.  Further examples
of band Chern insulators and $Z_2$ TIs are also provided in the SM.

Consider a gapped eigenstate of a many-body Hamiltonian in two
dimensions,
$|\Psi(\vec \kappa)\rangle = \sum_{i=1}^M \Psi_i(\vec \kappa)
|B_i\rangle$, where $\vec \kappa = (\kappa_x, \kappa_y)$ are twisted
boundary phases, $\kappa_{x,y} \in [0, 2\pi)$.  $\{|B_i\rangle\}$ are
orthonormal many-body bases independent of $\vec \kappa$, and
$\Psi_i(\vec \kappa) = \langle B_i | \Psi(\vec \kappa)\rangle$ is
periodic in $\vec \kappa$. The Chern number of $\Psi$ is
$C = \frac{1}{2\pi}\iint_{\text{BZ}} d^2 \kappa\, \nabla_{\vec \kappa}
\times \langle \Psi |i \nabla_{\vec \kappa}|\Psi\rangle.$ We first
show that $C$ can be computed using any two components of
$|\Psi(\vec \kappa)\rangle$, say $\Psi_{i_1}(\vec \kappa)$ and
$\Psi_{i_2}(\vec \kappa)$, provided they do not vanish at the same
$\vec \kappa$ point(s). We adopt the gauge fixing scheme of
Ref.~\cite{Kohmoto85}.  Assume for simplicity that a component
$\Psi_{i_1}(\vec \kappa)$ has a single zero in the entire BZ at, say,
$\vec \kappa^{*}$. Cases with multiple such zeros will be discussed
later. Divide the BZ into two patches, where one patch, denoted as
$R_2$, is an infinitesimal neighborhood around $\vec \kappa^{*}$, and
the remainder of the BZ is the other patch, denoted as $R_1$. We
choose the gauge of $|\Psi\rangle$ such that
\begin{gather}
  \label{gauge-12}
  \Psi_{i_a}(\vec \kappa) > 0 \text{ for } \vec \kappa \in R_a\ , \ a
  = 1,2\ .
\end{gather}
The gauge of $|\Psi\rangle$ is therefore smooth in both $R_1$ and
$R_2$, but has a phase mismatch across their interface,
\begin{gather}
  \label{mismatch-def}
  |\Psi(\vec \kappa_{\cap})\rangle_{R_1} = e^{i\lambda(\vec \kappa_{\cap})}|\Psi(\vec \kappa_{\cap})\rangle_{R_2}\ , \ \vec \kappa_{\cap} \in R_1 \cap R_2\ ,
\end{gather}
where subscripts $R_i$ denote gauge choice.  In gauge $R_1$, one can
write $(\Psi_{i_1}, \Psi_{i_2})_{R_1} = (r_1, r_2 e^{i\chi})$ with
$r_{1,2} > 0$ and real $\chi$. Then under gauge $R_2$,
$(\Psi_{i_1}, \Psi_{i_2})_{R_2} = (r_1 e^{-i\chi}, r_2)$. By
Eq.~\ref{mismatch-def}, one can identify $\lambda = \chi$,
\emph{viz.},
\begin{gather}
  \label{lambda-def}
  \lambda(\vec \kappa_{\cap}) = \Arg \left[ \Psi_{i_2}(\vec \kappa_{\cap}) / \Psi_{i_1}(\vec \kappa_{\cap})\right]\ ,
\end{gather}
which is \emph{gauge invariant}.  The BZ integral for computing $C$ is
now a sum over the two patches $R_{1,2}$, and by Stokes Theorem, each
patch contributes a line integral of the Berry connection vector over
the patch's boundary, thus
\begin{gather}
  \label{c-wind}
  C = \frac{1}{2\pi}\sum_{i=1,2} \oint\limits_{\partial R_i} d\vec
  \kappa_{\cap} \cdot \langle \Psi| i\nabla_{\vec \kappa_{\cap}} |
  \Psi\rangle_{R_i} = w[\lambda]\ ,
\end{gather}
where
$w[\lambda] = \frac{1}{2\pi} \ointctrclockwise_{\,\partial R_2}
d\vec \kappa_{\cap} \cdot \partial_{\vec \kappa_{\cap}} \lambda$ is
the winding number of the phase mismatch $\lambda(\vec \kappa_{\cap})$
in the counter-clockwise direction---note that the two boundaries,
$\partial R_1$ and $\partial R_2$, are identical but in opposite
directions. If $\Psi_{i_1}$ has multiple zeros, one can define a phase
mismatch $\lambda_a$ around the $a^{th}$ zero, and
$C = \sum_a w[\lambda_a]$. Eqs.~\ref{lambda-def} and \ref{c-wind}
together establish that the Chern number of $|\Psi\rangle$ can be
computed using any two of its components.

Now consider a truncated state $|\wt \Psi\rangle$ obtained by taking a
subset of wavefunction components from $|\Psi\rangle$ and
renormalizing. Its Chern number can be computed in the same way using
$\wt\Psi_{i_1}$ and $\wt\Psi_{i_2}$. Since both are simply rescaled
from their pre-truncation values, the phase mismatch
(Eq.~\ref{lambda-def}) is not affected by the truncation, hence
$|\wt\Psi\rangle$ and $|\Psi\rangle$ have the same Chern number.

\newcommand{\kiip}{\kappa_{\textsf{SIP}}}
\paragraph{Truncation invariance of quantized Berry phase}
Symmetry-protected 1D topological phases exhibit a robust $Z_2$ index
due to the quantization of the Berry phase to either $0$ or $\pi$. We
now prove the truncation invariance of the $Z_2$ class protected by
inversion-like symmetries. Examples in this class include the
Su-Schrieffer-Heeger model, Kitaev's $p$-wave superconductor, and
Spin-$1$ antiferromagnetic chain.
Consider a parent many-body Hamiltonian
$H(\kappa) = H(\kappa + 2\pi)$, where $\kappa \in [0, 2\pi)$ is the
boundary phase. Inversion-like invariance is defined as
$SH(\kappa) S^{-1} = H(-\kappa)$ where the unitary $S$ represents the
symmetry operation. At the symmetry invariant points
$\kiip \in \{0, \pi\}$, $S$ commutes with $H(\kiip)$, hence the ground
state of $H$, assumed unique, must also be a symmetry eigenstate,
$S|\Psi(\kiip)\rangle = s_{\kiip} |\Psi(\kiip)\rangle$, where
$s_{\kiip} = \pm 1$. Hughes \emph{et al} showed \cite{Hughes11} that
the Berry phase of $|\Psi(\kappa)\rangle$ can be computed from the
symmetry eigenvalues at $\kiip$, $e^{i\gamma} = s_0s_{\pi}$.
Now consider a truncation $P$ that preserves inversion, $[P,S] = 0$.
It follows that the truncated state $P|\Psi(\kiip)\rangle$ remains an
inversion eigenstate with the same eigenvalue $s_{\kiip}$ as the
parent state $|\Psi(\kiip)\rangle$. Hence, the Berry phase also
remains invariant, provided $P$ does not annihilate
$|\Psi(\kappa)\rangle$ for \emph{any} $\kappa$.

\paragraph{Parton construction of fractional Chern insulators}
Truncation invariance of the Chern number is closely related to the
parton construction of fractional Chern insulator (FCI) states
\cite{McGreevy12, Lu12, Zhang13, Hu15, Kourtis14}. Consider the
$\textsf{SU}(m)$ FCI state \cite{McGreevy12, Lu12}, a lattice analogue of the
Laughlin $\frac{1}{m}$ state. One writes the electron (or boson)
operator as a product of $m$ partons,
$c_{\vec r} = \prod_{\alpha = 1}^m f_{\vec r}^{(\alpha)}$. Each parton
species is subjected to a tight binding Hamiltonian with lowest band
Chern number $1$. Filling one band per species then leads to a parton
mean field state $|\Psi_{\text{MF}}\rangle$ with Chern number
$C_{\text{MF}} = m$ by construction. The FCI state is obtained by
Gutzwiller projecting $|\Psi_{\text{MF}}\rangle$ back to the electron
Hilbert space,
$|\Psi_{\text{el}}\rangle \propto P_G|\Psi_{\text{MF}}\rangle$, that
is, $0$ or $m$ partons per lattice site. From truncation invariance,
$|\Psi_{\text{el}}\rangle$ and $|\Psi_{\text{MF}}\rangle$ have the
same Chern number over a parton BZ, $\kappa_{x,y} \in [0,
2\pi)$. Here, $\kappa_{x,y}$ are parton twisted boundary phases,
$f_{\vec r + N_i \hat e_i}^{(\alpha)} = e^{i\kappa_i}
f^{(\alpha)}_{\vec r}$. The corresponding boundary conditions for
electrons are
$c_{\vec r + N_i \hat e_i} = \prod_{\alpha = 1}^m f_{\vec r + N_i \hat
  e_i}^{(\alpha)} = e^{im\kappa_i}c_{\vec r}$, hence one parton BZ is
equivalent to $m^2$ electron BZs. Thus although the Chern number
remains invariant after truncation when computed using the parton BZ,
the physical Hall conductance is related to the Chern number per
electron BZ \cite{Niu85}, and we recover the fractional Hall
conductance of $|\Psi_{\text{el}}\rangle$ as
$C = \frac{C_{\text{MF}}}{m^2} = \frac{1}{m}$.

\begin{figure}
  \centering
  \subfloat[Parton mean field state]{
    \includegraphics[width=.235\textwidth]{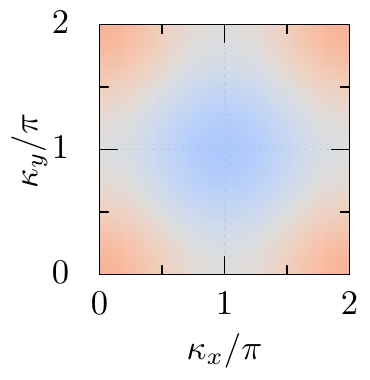}
  }
  \subfloat[Bosonic FCI state]{
    \includegraphics[width=.22\textwidth]{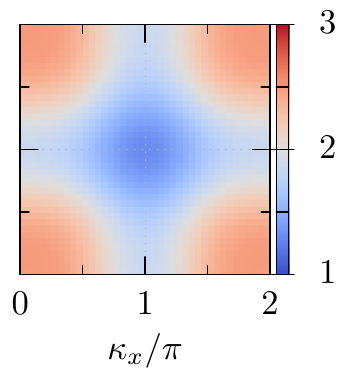}
  }
  \caption{Chern number density in the space of parton boundary phases
    for (a) the parent (untruncated) parton mean-field state, and (b)
    the bosonic fractional Chern insulator state obtained via
    Gutzwiller projection. In both cases, the Chern number density
    integrates to the same $C_{\text{MF}} = 2$ over the parton BZ, as
    required by truncation invariance. The physical Hall conductance
    of the FCI state is given by
    $C = \frac{C_{\text{MF}}}{m^2} = \frac{1}{2}$ with $m = 2$ parton
    species, see text for detail. Calculation is done with a
    $4\times 4$ lattice and a $40\times 40$ grid of
    $(\kappa_x, \kappa_y)$.}
  \label{fig:fci}
\end{figure}

In Fig.~\ref{fig:fci}, we use the $\pi$-flux square lattice model of
Ref.~\cite{Zhang13} as the mean field Hamiltonian for $m=2$ parton
species, and plot the Chern number density for both the untruncated
parton mean field state $|\Psi_{\text{MF}}\rangle$ and the bosonic FCI
state obtained by Gutzwiller projecting $|\Psi_{\text{MF}}\rangle$ to
$0$ or $2$ partons per site. In both cases, the Chern number density
integrates to $C_{\text{MF}}=2$ over the parton BZ, as guaranteed by
truncation invariance. The physical Hall conductance is given by
$C = \frac{C_{\text{MF}}}{m^2} = \frac{1}{2}$.

We note that numerical calculations of the fractional Chern number of
Gutzwiller-projected parton states are severely limited by system size
\cite{Hu15}. Our theorem establishes such results on a more general
ground, without system size restriction. The same argument applies to
the ground states of non-Abelian FCIs as well (see SM), although its
connection with quasi-particle statistics remains an open question.

\paragraph{Spin-$1$ antiferromagnetic chain}
We use the Spin-$1$ AKLT and Heisenberg models to illustrate
truncation invariance of the quantized Berry phase \cite{Haldane1983,
  Hatsugai2008}. The Hamiltonian is
$H(\kappa) = \sum_{i=1}^N \vec S_i \cdot \vec S_{i+1} + \beta (\vec
S_i \cdot \vec S_{i+1})^2$, where $\kappa$ is a boundary phase:
$S_{N+1}^{\pm} = S_1^{\pm}e^{\mp i\kappa}$ and $S_{N+1}^z =
S_1^z$. Define inversion $\mathcal{I}$ as
$\mathcal{I} \vec S_i \mathcal{I}^{-1} \equiv \vec S_{N+1-i}$, then
$H(\kappa)$ is inversion symmetric,
$\mathcal{I}H(\kappa) \mathcal{I}^{-1} = H(-\kappa)$. For
$|\beta| < 1$, its gapped ground state $|\Psi(\kappa)\rangle$ has a
nontrivial $Z_2$ index characterized by a quantized $\pi$ Berry
phase. We first consider the AKLT $\beta = \frac{1}{3}$, for which
$|\Psi(\kappa)\rangle$ can be obtained analytically \cite{AKLT87,
  Arovas88},
$|\Psi(\kappa)\rangle = \prod_{i=1}^N (a_i\dg b_{i+1}\dg - b_i\dg
a_{i+1}\dg)|\emptyset\rangle$, where $a$ and $b$ are Schwinger bosons,
$S_i^+ = a_i\dg b_i$, $S_i^z = \frac{1}{2}(a_i\dg a_i - b_i\dg b_i)$,
$a_i\dg a_i + b_i\dg b_i \stackrel{!}{=} 2$, $|\emptyset\rangle$ is
the boson vacuum, and $(a_{N+1}, b_{N+1}) = (a_1, b_1 e^{-i\kappa})$.
Now project $|\Psi(\kappa)\rangle$ onto two inversion conjugate spin
configurations $|B\rangle = |s_1^z, s_2^z, \cdots, s_N^z\rangle$ and
$|\bar B\rangle = \mathcal{I} |B\rangle$, $s_i^z \in \{0, \pm 1\}$.
To have $\langle B | \Psi(\kappa)\rangle \neq 0$, the nonzero spins in
$|B\rangle$ must have alternating signs, a manifestation of string
order \cite{Rommelse87, Girvin89}. One can show that the normalized
truncated wavefunction is
$|\wt \Psi(\kappa)\rangle = \frac{1}{\sqrt{2}}(|B\rangle + (-1)^N
e^{is\kappa}|\bar B\rangle)$, where $s$ is the leftmost nonzero spin
in configuration $|B\rangle$. This form is largely fixed by the
inversion conjugacy between $|B\rangle$ and $|\bar B\rangle$, which
ensures that (1) they have the same number of nonzero spins, and hence
are of equal absolute weight, and (2) their leftmost nonzero spins are
opposite, which leads to the relative phase $e^{is\kappa}$. Spoiling
either condition will lead to a non-quantized Berry phase. See SM for
derivation.  Parametrized on a Bloch sphere, $|\wt \Psi\rangle$ lies
on the equator and manifestly has a winding number $w_{\wt \Psi} = s$,
hence its Berry phase is $s\pi \equiv \pi \mod 2\pi$.

\begin{figure}
  \centering
    \hspace*{10pt}\includegraphics[width=0.47\textwidth]{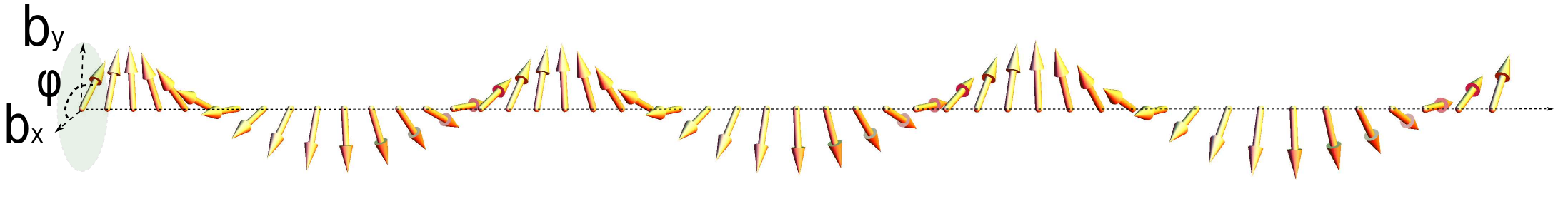}
    \includegraphics[width=0.48\textwidth]{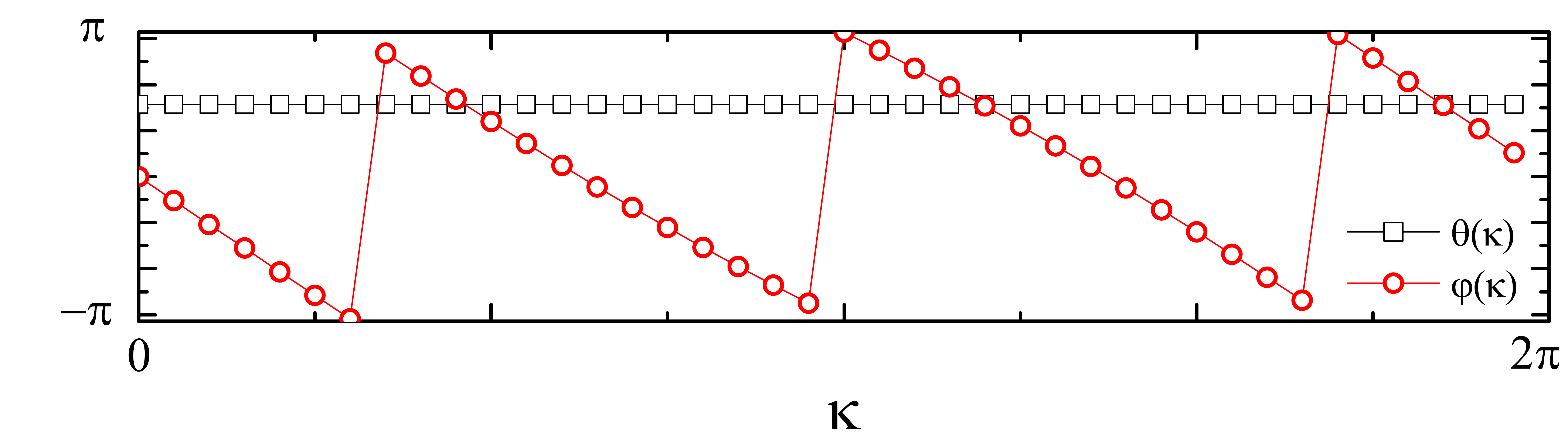}
    \caption{Projected Heisenberg ground state. (Top) Schematic plot
      showing the helical precession of the Bloch vector parametrizing
      the projected state,
      $|\wt \Psi\rangle = \cos \frac{\theta}{2} |B\rangle + \sin
      \frac{\theta}{2} e^{i\varphi} |\bar B\rangle$.  (Bottom)
      Spherical angles $\varphi$ and $\theta$. Over the cycle
      $\kappa = 0 \rightarrow 2\pi$, $\theta$ remains a constant
      $\frac{\pi}{2}$, and $\varphi(\kappa)$ changes by $-6\pi$, hence
      the winding number of $\hat b$ is $-3$, consistent with a Berry
      phase of $\pi(\text{mod } 2\pi)$. $N=12$ spin-1 sites are used.
      The ground state is truncated to the many-body basis
      $|B\rangle = |\ua\ua\ua\ua 0\da\da\da\da\ua\da 0 \rangle$ and
      its inversion partner.  }
 \label{fig:s1chain}
\end{figure}

When $\beta \neq \frac{1}{3}$, the Hamiltonian is no longer a
projection operator onto bond singlets, hence there is a proliferation
of spin configurations in the ground state that violate the
sign-alternating string order, and the winding number of a truncated
state, $w_{\wt\Psi}$, is not restricted to $\pm 1$. Nevertheless,
since inversion symmetry is intact, the post-truncation Berry phase
remains $\pi$, indicating that $w_{\wt \Psi}$ is an odd integer. Using
the Heisenberg model ($\beta = 0$), we have numerically verified that
(1) if $|B\rangle$ and $|\bar B\rangle$ are string ordered, the
winding number remains $\pm 1$; if not, the winding number is an odd
integer but not necessarily $\pm 1$, see Fig.~\ref{fig:s1chain}. (2)
If we instead twist the Hamiltonian on the bond between $\vec S_\ell$
and $\vec S_{\ell+1}$, the new winding number $w_{\wt \Psi}^{(\ell)}$
is related to $w_{\wt \Psi}$ via a ``Gauss law'',
$w_{\wt \Psi}^{(\ell)} - w_{\wt\Psi} = - 2 \sum_{n=1}^{\ell} s_n^z$,
suggesting that $s_n^z = \pm 1$ in a given spin configuration act as
charge $\mp 2$ sources of winding numbers. (3) For projections that
violate inversion symmetry, the Berry phase is in general not
quantized any more. These results are numerically robust even though
the typical weight on a many-body basis state is exponentially small
($\sim \frac{1}{\sqrt{3^N}}$).

\paragraph{Relation with bulk entanglement}
Connection between Hilbert space truncation and topology has
previously been studied from the perspective of quantum entanglement
\cite{Peschel03, Cheong04, Li08, Thomale10, Pollmann10b, Prodan10,
  Hughes11, Huang12}.  We briefly discuss the relation between
entanglement and wavefunction truncation in the context of bulk
entanglement \cite{Hsieh14, Fukui14, Chiou16, Legner13, Schliemann13}
due to a sublattice bipartition.  Consider a single occupied Bloch
band $|\psi_{\vec k}\rangle$ with momentum $\vec k$. Generalization to
multiple occupied bands is straightforward. The Schmidt decomposition
of $|\psi_{\vec k}\rangle$ into two sublattice groups $A$ and $B$ is
\begin{gather}
  \label{schmidt}
  |\psi_{\vec k}\rangle = \sqrt{f_{\vec k}} |\wt\psi_{A,\vec k}\rangle
  \otimes |\emptyset_B\rangle + \sqrt{1-f_{\vec k}}
  |\emptyset_A\rangle \otimes |\wt\psi_{B,\vec k}\rangle,
\end{gather}
where $|\emptyset_{A(B)}\rangle$ and $|\wt\psi_{\vec k,A(B)}\rangle$
are respectively the vacuum and the truncated state in part $A(B)$,
$f_{\vec k} = \langle \psi_{\vec k}|P_A |\psi_{\vec k}\rangle$.
$|\wt\psi_{\vec k,A}\rangle$ is thus an entanglement eigenstate for
part $A$ in the single particle sector, with entanglement eigenvalue
$f_{\vec k}$. For a partition with $N_A$ sublattices in $A$, there
should be a total of $N_A$ (single particle) entanglement levels,
thus $N_A - 1$ of them are identically zero. If
$f_{\vec k} \neq 0 \forall \vec k$, it is gapped from the remainder,
hence one can introduce a topological index, such as an entanglement
Chern number \cite{Fukui14}, for the corresponding entanglement
eigenstate, \emph{i.e.}, the truncated state
$|\wt\psi_{\vec k, A}\rangle$. Truncation invariance thus implies that
the \textit{entanglement topological index} must be identical to the
topological index of the parent state if (1) the bulk entanglement
spectrum is gapped from zero, and (2) for SPT parent states, the
entanglement partition preserves the protecting symmetry.

\paragraph{Measuring topological index via partial tomography}
Truncation invariance of the topological index is experimentally
relevant. Recent breakthrough in quench-based quantum tomography has
made it possible to extract topological indices of
\emph{two}-component Bloch wavefunctions by performing a full
measurement of both wavefunction components over the entire BZ (of
Bloch momenta) \cite{Hauke14, Flaschner16a}.  We now discuss a
quench-based partial quantum tomography for a multi-component Bloch
wavefunction
$|\psi(\vec k)\rangle = \sum_{a=1}^N \psi_a(\vec k) |a\rangle$, from
which two chosen components $\psi_{a_1}(\vec k)$ and
$\psi_{a_2}(\vec k)$ can be measured. Here $a$ labels sublattices
within a unit cell. Combined with truncation invariance, this allows
us to determine the Chern number of the full state
$|\psi(\vec k)\rangle$. We follow the experimental protocol of
Refs.~\cite{Hauke14, Flaschner16a}. Assume at $t=0$ the system has
been prepared as a filled Bloch band described by
$|\psi(\vec k)\rangle$. For $0 < t < t_h$, we quench the system with a
flat band Hamiltonian
$H(\vec k) = \sum_{a=1}^N\varepsilon_a |a\rangle\langle a|$. The
values of $\{\varepsilon_a\}$ will be specified later. At the end of
the quench, one has
$|\psi(\vec k, t_h)\rangle = \sum_{a=1}^N\psi_a(\vec k,t_h) |a\rangle$
where $\psi_a(\vec k, t_h) = \psi_a(\vec k) e^{-i\varepsilon_a
  t_h}$. The system is then released for a time of flight (TOF)
measurement. The resulting momentum distribution from the TOF analysis
is given by \cite{Hauke14}
$n(\vec k,t_h) = \left|\sum_{a=1}^N \psi_a(\vec k, t_h)\right|^2$, and
by monitoring $n(\vec k, t_h)$ as a continuous function of $t_h$,
contributions from different $\psi_a(\vec k)$ (at $t=0$) can in
principle be resolved.

To perform a partial tomography on, say, the first two sublattices
$a = 1, 2$, we set $\varepsilon_a$ for all other sublattices $a>2$ to
a common level $E$, and require that
$\varepsilon_1 \neq \varepsilon_2 \neq E$. Consequently, the momentum
distribution $n(\vec k, t_h)$ has only three distinctive frequency
modes,
\begin{gather}
  \omega_{1(2)} = \varepsilon_{1(2)} - E\ , \ \omega_3 = \varepsilon_2
  - \varepsilon_1\ ,
\end{gather}
and from the TOF experiment, one can extract the corresponding Fourier
coefficients $A_i, B_i$,
\begin{gather}
  n(\vec k, t_h) = A_0(\vec k) + \sum_{i=1}^3 \Bigl[A_i(\vec
  k)\cos(\omega_i t_h) + B_i(\vec k) \sin(\omega_i t_h) \Bigr]\ .
\end{gather}
Parametrize $\psi_1 = u \sin \frac{\theta}{2}$ and
$\psi_2 = -u \cos \frac{\theta}{2}e^{i\varphi}$, $u>0$. The overall
scale $u$ does not enter the topological index evaluation. The Bloch
vector angles $\varphi$ and $\theta$ are
\begin{gather}
  \label{phi-theta-tomo}
  \tan \varphi(\vec k) = \frac{B_3(\vec k)}{A_3(\vec k)}\ , \
  \tan \frac{\theta(\vec k)}{2} = \sqrt{\frac{A_1^2(\vec k) + B_1^2(\vec k)}{A_2^2(\vec k) + B_2^2(\vec k)}}\ ,
\end{gather}
see SM for derivation
and $\varphi$,$\theta$ plots of a truncated Hofstadter
band. Eq.~\ref{phi-theta-tomo} allows us to extract the projected
state $|\wt\psi\rangle = (\frac{\psi_1}{u}, \frac{\psi_2}{u})^t$, from
which the Chern number of the full state can be computed.  In fact,
since $|\wt\psi\rangle$ is also a bulk entanglement eigenstate, this
is a measurement protocol for the entanglement Chern number of a
sublattice truncation as discussed in the previous section.

\paragraph{Conclusion}
We have shown that a normalizable truncated wavefunction preserves the
topological index of its parent state, if both indices are computed in
the space of the parent state's twisted boundary phases. The physical
interpretation of the index may change for the truncated state if its
boundary condition is affected by the truncation, and we gave an
example using the parton construction of the $\textsf{SU}(m)$ FCI state.  We
also showed that a sublattice-truncated state can be identified as an
entanglement eigenstate resulting from a sublattice bipartition,
revealing a connection between wavefunction truncation and quantum
entanglement.
Our finding provides a new perspective on the topological structure of
wavefunctions, %
and indicates that mathematical specification of a topological index,
and perhaps even its physical manifestation, can be achieved in a much
smaller Hilbert space, such as the 2-sublattice space that may be
probed by the partial tomography scheme discussed in the text.

\begin{acknowledgments}
  \paragraph{Acknowledgments}%
  We are grateful to Yi Zhang for critical reading and comments of an
  early draft, and to D.~N.~Sheng, Kai Sun, Christof Weitenberg, Avadh
  Saxena, Hongchul Choi, and S.~Kourtis for useful discussions and
  communications.  W.Z. thanks T.~S. Zeng for helpful discussion and
  F.~D.~M. Haldane for education of the physics of spin-1
  antiferromagnetic chain.  Work at LANL was supported by US DOE BES
  E3B7 (ZSH, JXZ, and AVB), and by US DOE NNSA through LANL LDRD (ZSH,
  WZ, and AVB). Work at NORDITA was supported by ERC DM 321031 (AVB).
\end{acknowledgments}

\bibliography{trinv} \bibliographystyle{apsrev-no-url}

\onecolumngrid
\section{Supplemental Materials}

In this note, we give additional examples and derivations showing
truncation invariance in (1) the Hofstadter model, a band Chern
insulator, (2) the BHZ model, a time-reversal-invariant $Z_2$
topological insulator, (3) the AKLT model, which belongs to the $Z_2$
Haldane phase and (4) a fractional Chern insulator model hosting
non-Abelian fractional quantum Hall effect.  We also provide
derivation details of the partial quantum tomography scheme introduced
in the main text, and discuss the relation and distinction of
truncation invariance with recent works on the node structure in
wavefunction overlaps.

\section{Hofstadter model}
We go through the general proof of truncation invariance of the Chern
number in detail, and and provide additional demonstrations, using the
paradigmatic Hofstadter model \cite{Hofstadter76}. This model
describes electrons hopping on a square lattice in the $xy$ plane
placed in a uniform magnetic field along $z$. For a rational flux per
square plaquette, $\phi = 2\pi p/q$ ($p$ and $q$ are coprime
integers), the magnetic unit cell consists of $q$ consecutive
plaquettes, which we choose to align in the $y$
direction. Correspondingly, there are $q$ Bloch bands. Each band
wavefunction can be expressed as a $q$-element column vector,
$|\psi(\vec k)\rangle = (\psi_1(\vec k), \psi_2(\vec k), \cdots,
\psi_q(\vec k))^t$, where
$\psi_a(\vec k) = \langle a | \psi(\vec k)\rangle$ and $|a\rangle$ is
the atomic state on the $a^{th}$ site of the magnetic unit cell.

\subsection{Wavefunction zeros and phase vortices}
\begin{figure}
  \centering
  \includegraphics[width=.4\textwidth]{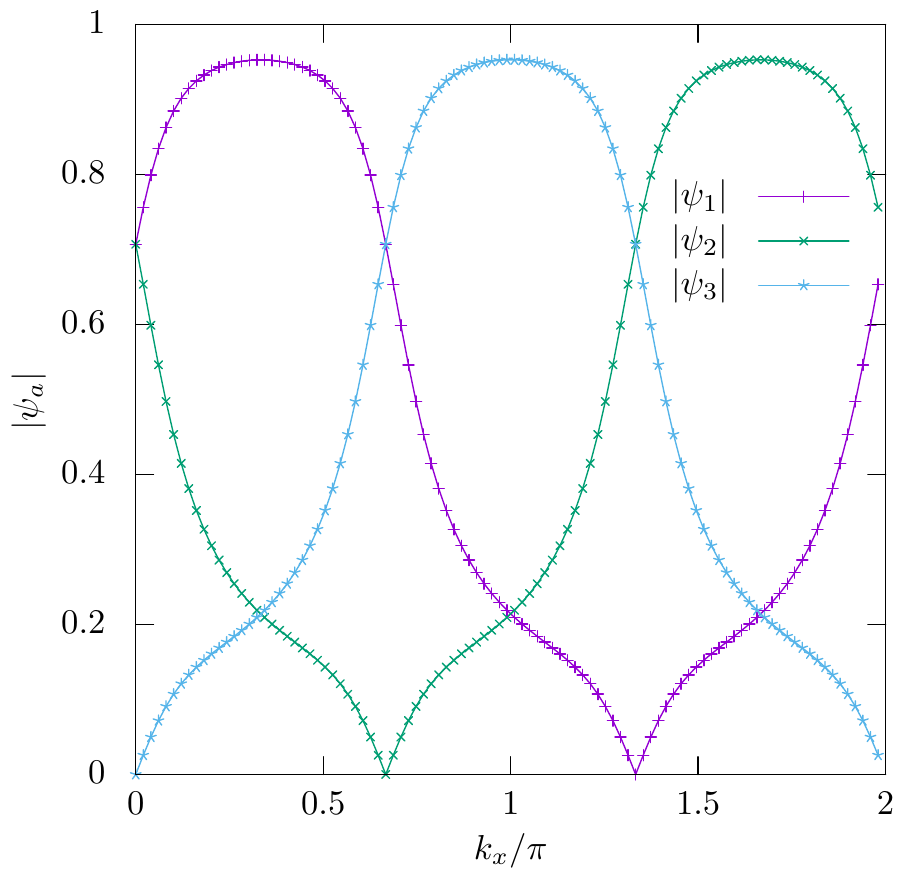}
  \caption{The lowest subband wavefunction of the $p/q = 1/3$
    Hofstadter model, at $k_y = 0$. All three components show one zero
    as $k_x$ varies from $0$ to $2\pi$, consistent with the band Chern
    number being $C = 1$. We have verified that the zeros only occur
    at $k_y = 0$.}
  \label{fig:hof-wf-nodes}
\end{figure}

In this section, we go through the general proof of truncation
invariance of the Chern number in more detail, using the three-band
case $p/q = 1/3$ as an example. The lowest band $|\psi(\vec k)\rangle$
has a Chern number $C=1$, therefore all three of its wavefunction
components, $\psi_i(\vec k), i = 1, 2, 3$, have at least one zero in
the Brillouin zone. One can verify that the zeros of $\psi_1, \psi_2$,
and $\psi_3$ occur at $k_y = 0$ and
$k_x = \frac{4\pi}{3}, \frac{2\pi}{3}$, and $0$, respectively, see
Fig.~\ref{fig:hof-wf-nodes}.

\begin{figure}
  \centering
  \includegraphics[width=.4\textwidth]{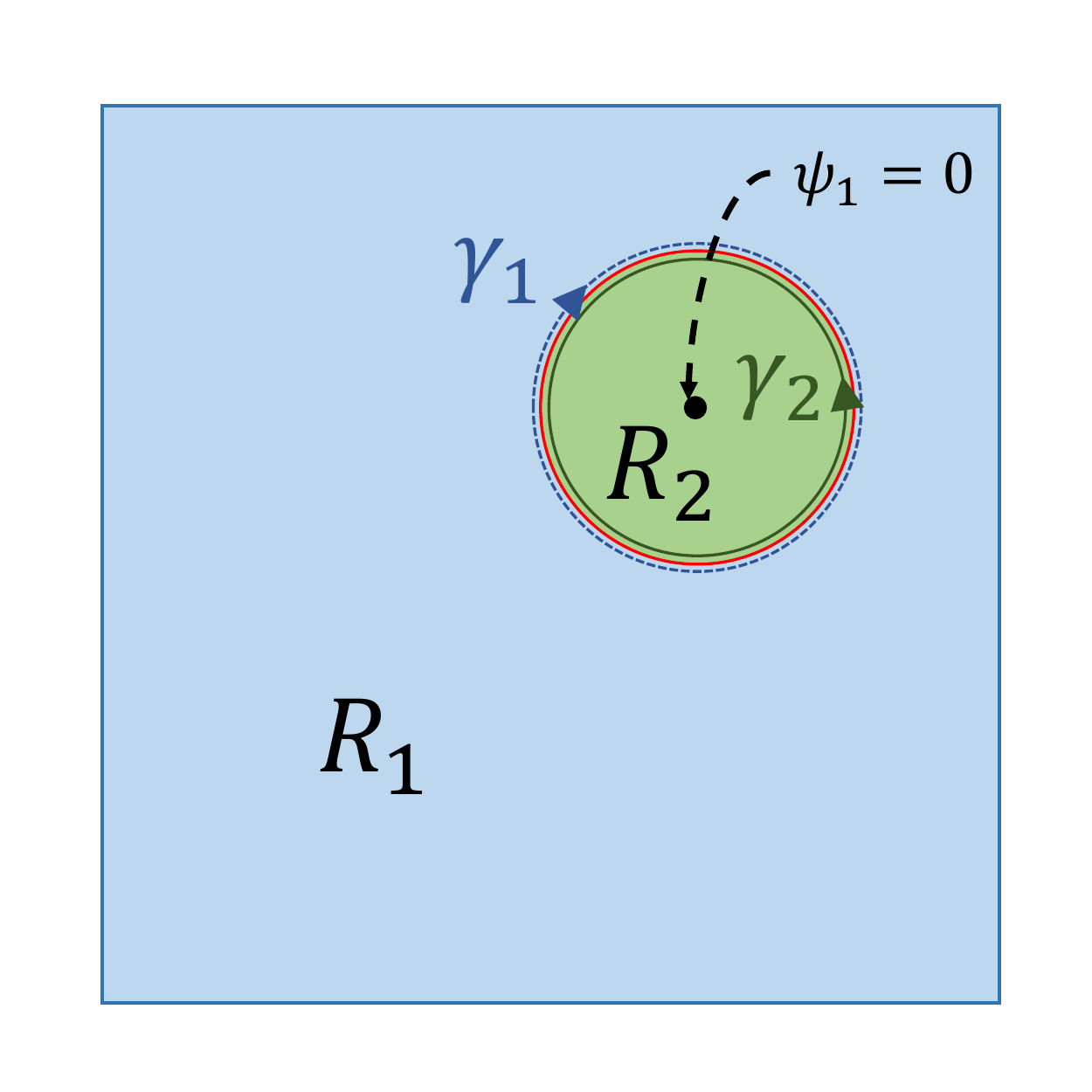}
  \caption{Schematics of the wavefunction gauge choice according to
    the zero of $\psi_1(\vec k)$. The Chern number computed with this
    gauge is the sum of two Berry phases, $\gamma_1$ along
    $\partial R_1$ in clockwise direction, and $\gamma_2$ along
    $\partial R_2$ in counter-clockwise direction.}
  \label{fig:hof-gauge}
\end{figure}

To compute the Chern number of $|\psi(\vec k)\rangle$, we now divide
the Brillouin zone into two patches, see Fig.~\ref{fig:hof-gauge}. One
patch, denoted as $R_2$, is an infinitesimal neighborhood of radius
$\epsilon$ around the zero of $\psi_1$, at
$\vec k_1 = (\frac{4\pi}{3}, 0)$:
$R_2 = \{ \vec k: |\vec k - \vec k_1 | \le \epsilon\}$. The remainder
constitutes the other patch,
$R_1 = \{ \vec k: |\vec k - \vec k_1| \ge \epsilon\}$. Since $\psi_1$
has only one zero at $\vec k_1$, one can always choose a gauge for
$R_1$ such that $\psi_1$ is real and positive,
\begin{gather}
  \label{psi-r1}
  |\psi(\vec k)\rangle_{R_1} =
  \begin{pmatrix}
    |\psi_1(\vec k)| \\ |\psi_2(\vec k)| e^{i\phi_2(\vec k)} \\ |\psi_3(\vec k)| e^{i\phi_3(\vec k)}
  \end{pmatrix}\quad, \quad \vec k \in R_1\ .
\end{gather}
We have used the subscript $R_1$ to denote the gauge choice. In the
patch $R_2$, we instead choose a gauge where $\psi_2$ is real and
positive. This is always achievable because the zeros of $\psi_1$ and
$\psi_2$ do not coincide, see Fig.~\ref{fig:hof-wf-nodes}. Thus
\begin{gather}
  \label{psi-r2}
  |\psi(\vec k)\rangle_{R_2} =
  \begin{pmatrix}
    |\psi_1(\vec k) | e^{i\varphi_1(\vec k)} \\
    |\psi_2(\vec k)|\\
    |\psi_3(\vec k)| e^{i\varphi_3(\vec k)}
  \end{pmatrix}\quad , \quad \vec k \in R_2\ .
\end{gather}
On the interface between the two patches, defined as
\begin{gather}
  \label{r1-r2}
R_1 \cap R_2 = \{\vec k_{\cap}: |\vec k - \vec k_1| = \epsilon\}\ ,  
\end{gather}
$|\psi(\vec k)\rangle_{R_1}$ and $|\psi(\vec k)\rangle_{R_2}$ differ
by an overall phase $\lambda(\vec k)$,
\begin{gather}
  |\psi(\vec k_{\cap})\rangle_{R_1} = e^{i\lambda(\vec k_{\cap})} |\psi(\vec
  k_{\cap})\rangle_{R_2}\quad , \quad \vec k_{\cap} \in R_1 \cap R_2\ .
\end{gather}
From Eqs.~\ref{psi-r1} and \ref{psi-r2}, one has that
\begin{align}
  \notag
  \lambda(\vec k_{\cap}) &= - \varphi_1(\vec k_{\cap}) = \phi_2(\vec k_{\cap})\\
  &= \Arg \frac{\psi_2(\vec k_{\cap})}{\psi_1(\vec k_{\cap})}\quad , \quad \vec k_{\cap} \in R_1 \cap R_2\ .
\end{align}
The second line is manifestly gauge invariant.

\begin{figure}
  \centering
  \subfloat[$\Arg \frac{\psi_2}{\psi_1}$]{
    \includegraphics[width=.5\textwidth]{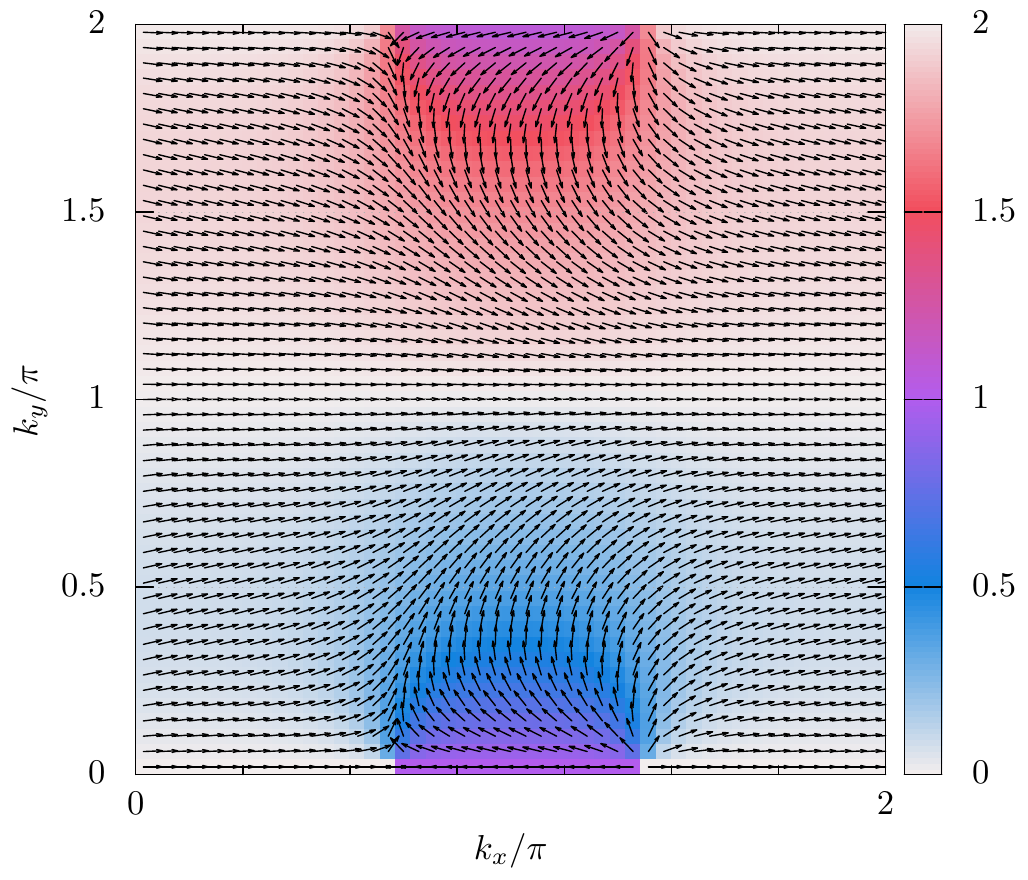}
  }
  \subfloat[$\Arg \frac{\psi_3}{\psi_2}$]{
   \includegraphics[width=.5\textwidth]{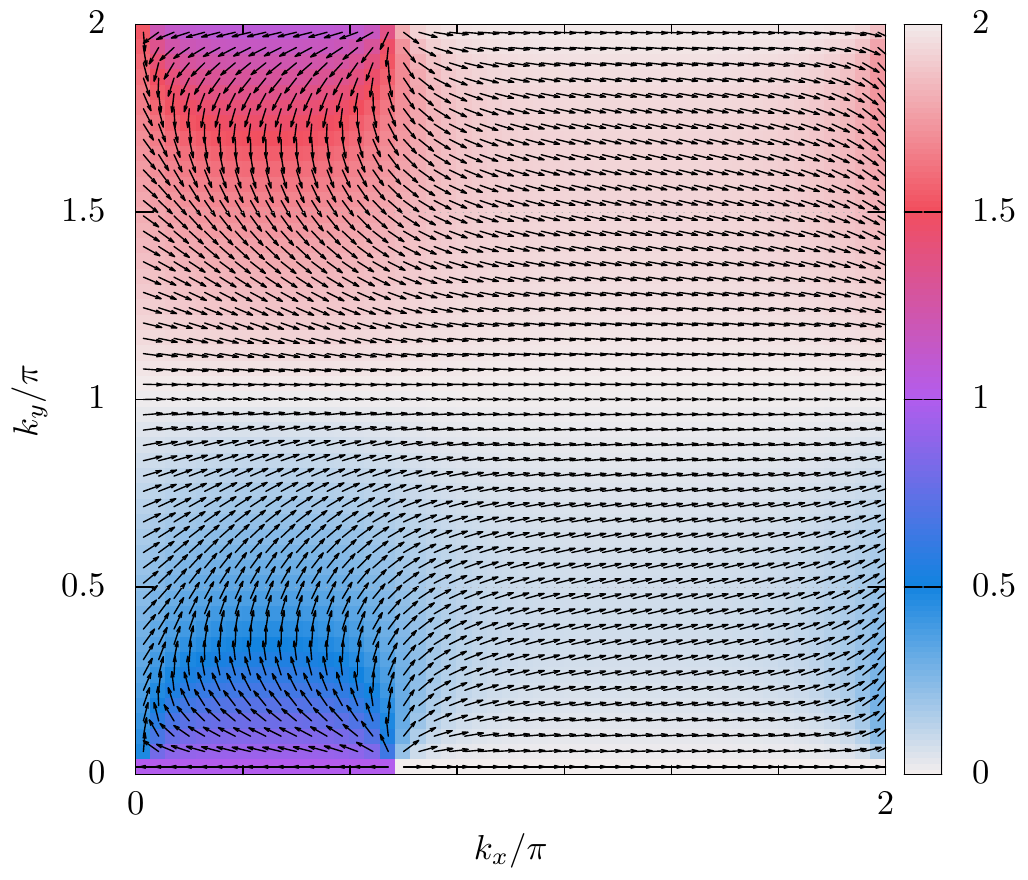}
  }
  \caption{Phases of $\psi_2/\psi_1$ and $\psi_3/\psi_2$ for the
    lowest band of the $p/q = 1/3$ Hofstadter model. Color encodes the
    phase angles in unit of $\pi$, and arrows show the same
    information represented as polar vectors
    $(\cos \varphi_{\vec k}, \sin \varphi_{\vec k})$. The phases show
    two singularities corresponding to the zero of the numerator and
    the denominator, respectively. The zero of the denominator is a
    vortex of the relative phase, around which the relative phase has
    a winding number $1$. The zero of the numerator is an anti-vortex
    of the relative phase, around which the relative phase has a
    winding number $-1$; see Fig.~\ref{fig:hof-wf-nodes} for the
    location of zeros. Following the general construction outlined in
    the text, the Chern number of both the truncated state and the
    untruncated state can be identified as the winding number of the
    phase around the zero of the denominator, and equivalently the
    negative winding number of around the zero of the numerator. }
  \label{fig:hof-phase-winding}
\end{figure}

The Chern number of $|\psi(\vec k)\rangle$ can now be computed as
\begin{gather}
  C = \frac{1}{2\pi} \iint\limits_{BZ} d^2k \nabla_{\vec k} \times
  \langle \psi(\vec k) | i\nabla_{\vec k}|\psi(\vec k)\rangle =
  \frac{1}{2\pi} (\iint\limits_{R_1} + \iint\limits_{R_2}) \cdots =
  \frac{1}{2\pi} \sum_{i = 1,2}\oint\limits_{\partial R_i} d\vec k_{\cap}
  \cdot \langle \psi(\vec k_{\cap}) | i\nabla_{\vec k_{\cap}} | \psi(\vec
  k_{\cap})\rangle_{R_i}\ ,
\end{gather}
where we have used Stokes theorem to convert the area integrals over
$R_1$ and $R_2$ into line integrals over their boundaries. Note that
the two boundaries, $\partial R_1$ and $\partial R_2$, are identical
but in opposite directions; both consist of the infinitesimal loop
Eq.~\ref{r1-r2}, with $\partial R_2$ in the counter-clockwise and
$\partial R_1$ in the clockwise direction. The Chern number is thus
\begin{gather}
  C = \frac{1}{2\pi} \ointctrclockwise\limits_{R_1\cap R_2} d\vec k_{\cap} \cdot \Bigl[\langle \psi(\vec k_{\cap}) | i\nabla_{\vec k_{\cap}} | \psi(\vec k_{\cap})\rangle_{R_2} - \langle \psi(\vec k_{\cap}) | i\nabla_{\vec k_{\cap}} | \psi(\vec k_{\cap})\rangle_{R_1}  \Bigr] = \frac{1}{2\pi} \ointctrclockwise\limits_{R_1 \cap R_2} d\vec k_{\cap} \cdot \nabla_{\vec k_{\cap}} \lambda(\vec k_{\cap}) \equiv w[\lambda]\ ,
\end{gather}
which is the counter-clockwise winding number of the gauge invariant
phase mismatch $\lambda(\vec k_{\cap})$. To obtain the second
equality, we have used
$\langle e^{-i\lambda(\vec k)}\psi(\vec k) | i\nabla_{\vec k}
|e^{i\lambda(\vec k)}\psi(\vec k)\rangle = \langle \psi(\vec k) |
i\nabla_{\vec k} | \psi(\vec k)\rangle - \nabla_{\vec k}\lambda(\vec
k)$. This is also equivalent to the difference of Berry phases
evaluated with the two different gauges $|\psi\rangle_{R_1}$ and
$|\psi\rangle_{R_2}$, over the same path $R_1 \cap R_2$ in
counter-clockwise direction, see Fig.~\ref{fig:hof-gauge}. It is known
that when evaluated with different gauge choices, the physical (gauge
invariant) Berry phase is only defined up to integer multiples $2\pi$,
and we see that the said integer, in this context, is the Chern number.

The above computational scheme for the Chern number can be summarized
as: The Chern number of $|\psi(\vec k)\rangle$ can be computed from
any two components, $\psi_{i_1}$ and $\psi_{i_2}$, as the winding
number of the \emph{gauge invariant} relative phase
$\Arg \frac{\psi_{i_2}}{\psi_{i_1}}$ around the zero of the
\emph{denominator} $\psi_{i_1}$. If multiple zeros exist, the Chern
number is the total vorticity around these zeros.

Since truncation does not change the ratio between any pair of
wavefunction elements, the Chern number of a renormalizable truncated
state must be the same as the parent state.

\subsection{Sublattice truncation invariance}
\begin{figure}[h]
  \centering
  \subfloat[$\cos\theta(\vec k)$]{
    \includegraphics[width=.32\textwidth, trim={0 0 10 0}, clip]{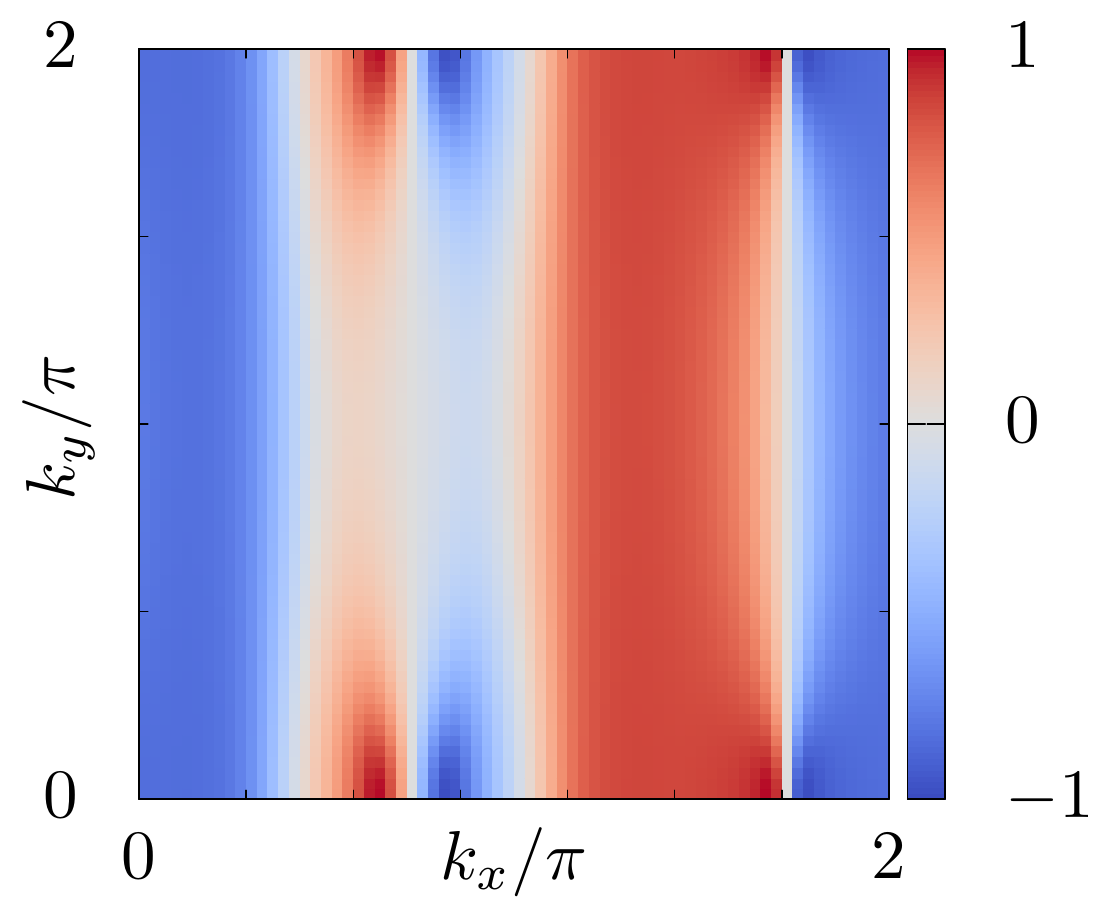}
  }
  \subfloat[$\varphi(\vec k)/\pi$]{
    \includegraphics[width=.275\textwidth, trim={35 0 0 0}, clip]{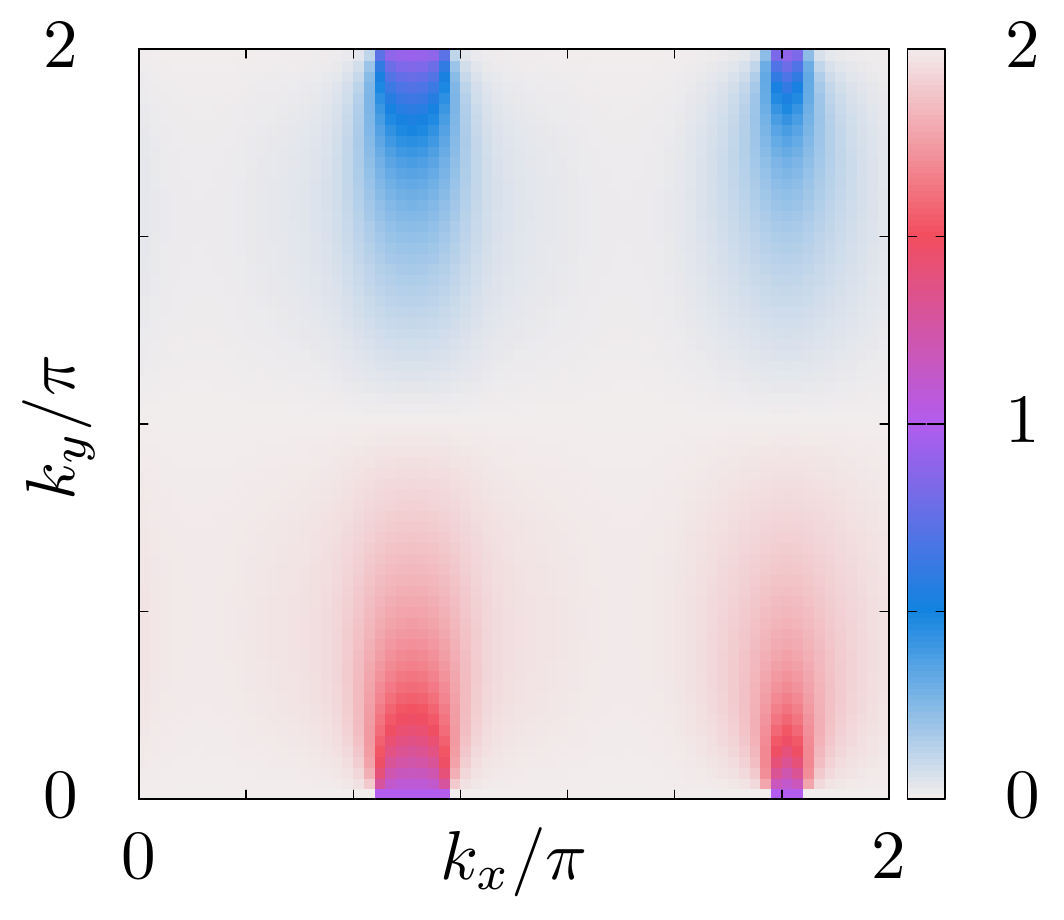}
  }
  \caption{Bloch vector representation of a $2$-element truncation of
    a Hofstadter band. We truncate the lowest band
    $|\psi(\vec k)\rangle$ to sublattices $1$ and $2$,
    $P|\psi\rangle/\sqrt{\langle \psi | P | \psi\rangle} = (\wt\psi_1,
    \wt\psi_2)^t \equiv (\sin \frac{\theta}{2}, -\cos
    \frac{\theta}{2}e^{i\varphi})$. The Hofstadter flux is set as
    $p/q = 3/7$, hence $|\psi(\vec k)\rangle$ has a Chern number
    $C=-2$. Panel (a): $z$-component of the Bloch vector,
    $b_z = \cos \theta$. Panel (b): its azimuthal angle $\varphi$. The
    north pole of the Bloch sphere ($b_z = 1$) can be identified as
    the darkest red spots in (a), and the south pole ($b_z = -1$) the
    darkest blue spots, all located on $k_y = 0$. Each pole is covered
    twice, consistent with $C = -2$. The azimuth $\varphi$ around each
    pole exhibits a vortex ($\varphi$ traverses $0 \rightarrow 2\pi$
    going around a pole), as can be verified in (b). Vertical white
    stripes in (a) correspond to the equator of the Bloch sphere
    ($b_z=0$); of the four such stripes, only two have $\varphi$
    winding from $0$ to $2\pi$ in (b), hence the equator is also
    covered twice, consistent with $C = -2$.}
  \label{fig:hof}
\end{figure}

We consider truncation to a $2$-sublattice Hilbert space,
$|\wt\psi(\vec k)\rangle \equiv (\wt\psi_{i_1}(\vec k),
\wt\psi_{i_2}(\vec k))^t$, where
$\wt\psi_{i_1}/\wt\psi_{i_2} = \psi_{i_1}/\psi_{i_2}$ and
$\langle \wt \psi | \wt\psi\rangle = 1$. $|\wt\psi(\vec k)\rangle$ can
be parametrized by a vector $\op b \equiv (\theta, \varphi)$ on the
unit Bloch sphere, $\wt \psi_{i_1} = \sin \frac{\theta}{2}$ and
$\wt \psi_{i_2} = -\cos \frac{\theta}{2} e^{i\varphi}$. This
parametrization is also used in the partial tomography discussed in
the text and a later section in this SM. The Chern number of $\wt\psi$
measures the number of times $\hat b$ covers the Bloch sphere,
$\wt C = \frac{1}{4\pi}\iint d^2k\ \op b(\vec k) \cdot
\Bigl[\partial_{k_x}\op b(\vec k) \times \partial_{k_y} \op b(\vec
k)\Bigr]$. In Fig.~\ref{fig:hof}, we plot the Bloch vector
$\op b(\vec k)$ for the state truncated to sublattices
$(i_1, i_2) = (1,2)$. The parent state is chosen as the lowest
Hofstadter band with flux $p/q = 3/7$, which has a Chern number of
$C=-2$. One can verify from Fig.~\ref{fig:hof} that $\wt C = C$.

\section{BHZ model}
We use the BHZ model to illustrate truncation invariance of the $Z_2$
class in 2D, which, in principle, follows from the invariance of the
spin Chern number. The BHZ model has a four-element unit cell
$(A_{\ua}, B_{\ua}, A_{\da}, B_{\da})$, where $A, B$ denote
sublattices and $\ua,\da$ denote spin. The Hamiltonian is \cite{BHZ06}
\begin{gather}
  \label{h-bhz}
  H(\vec k) = \sin k_x \sigma_z\otimes \tau_x + \sin k_y \id \otimes
  \tau_y + (2-m-\cos k_x - \cos k_y)\id \otimes \tau_z + \Delta
  \sigma_y\otimes\tau_y\ ,
\end{gather}
where $\tau$ and $\sigma$ are Pauli matrices acting on the sublattice
and spin spaces, respectively, and $\Delta \neq 0$ breaks inversion
symmetry.

We will implement a truncation by projecting out every other $A$ site
along the $x$ direction for both spin species. This effectively
doubles the unit cell along $x$, yielding an $8$-band model prior to
truncation. The Hamiltonian with doubled unit cell is
\begin{gather}
  H(q_x, k_y) =
  \begin{pmatrix}
    H_0 & H_1 + H_{-1} e^{-iq_x} \\
    H_{-1} + H_1 e^{iq_x} & H_0
  \end{pmatrix}\ ,
\end{gather}
where $q_x$ is the Bloch momentum with respect to the doubled unit
cell along $x$, and the $k_y$-dependent $4\times 4$ blocks are
\begin{gather}
  H_0(k_y) = \sin k_y \id \otimes \tau_y + (2-m-\cos k_y) \id \otimes \tau_z + \Delta \sigma_y \otimes \tau_y\ , \\
  H_{\pm 1}(k_y) = \pm \frac{1}{2i} \sigma_z \otimes \tau_x -
  \frac{1}{2} \id \otimes \tau_z\ .
\end{gather}

\begin{figure}
  \centering
  \subfloat[$Z_2 = 1$]{
    \includegraphics[width=.3\textwidth, trim={12 0 10 0}, clip]{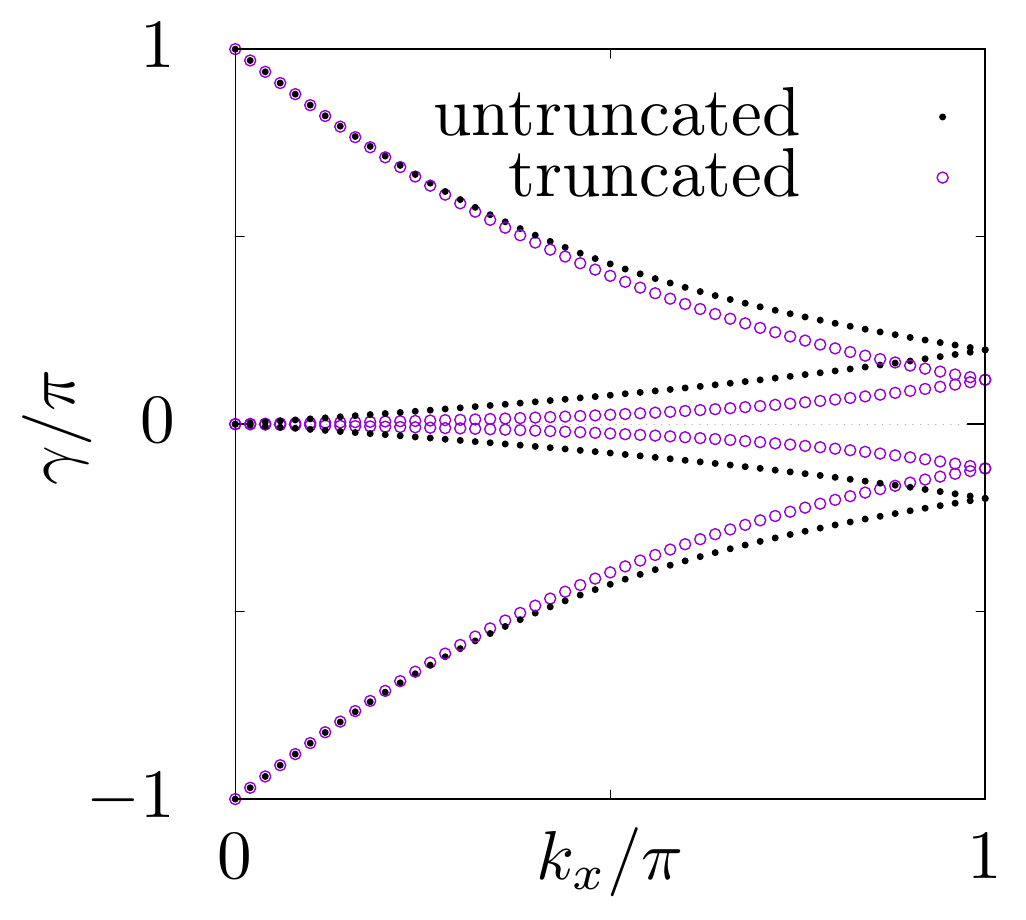}
  }
  \subfloat[$Z_2 = 0$]{
    \includegraphics[width=.305\textwidth, trim={5 0 10 0}, clip]{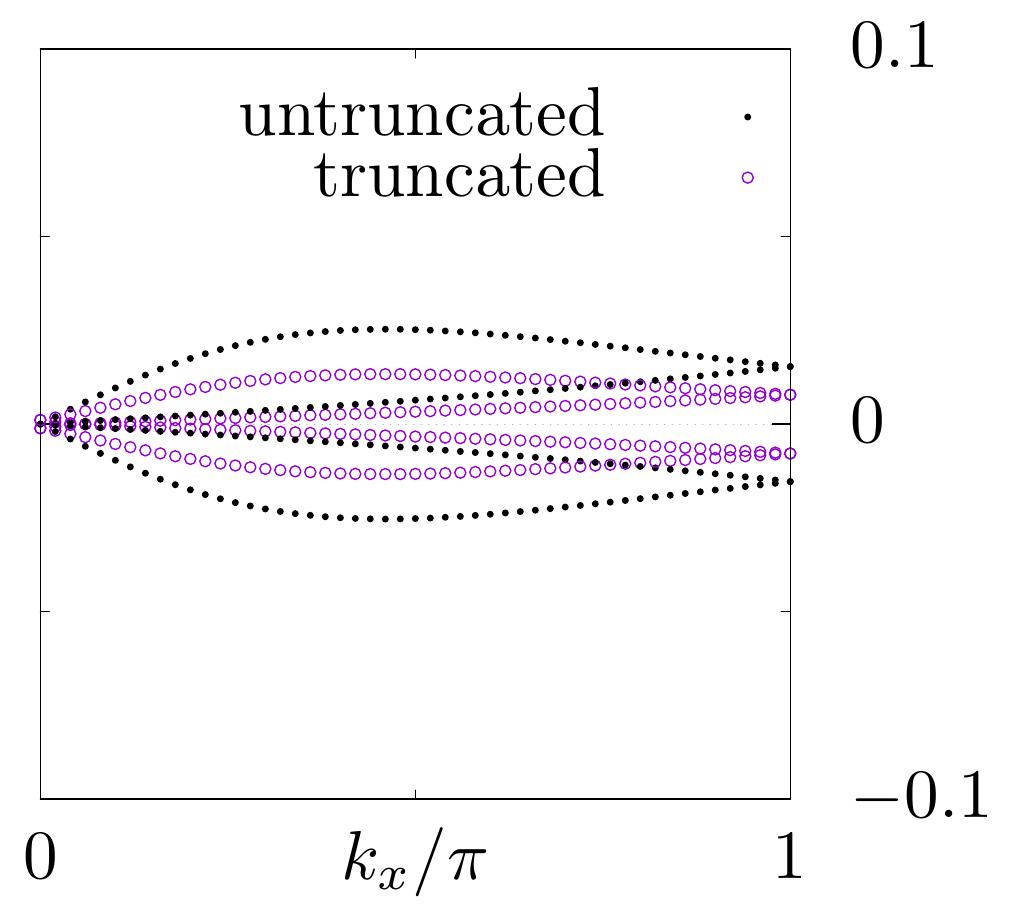}
  }
  \caption{Wannier spectral flow of the full and truncated BHZ ground
    state (at half filling), demonstrating that a symmetry-preserving
    truncation does not change the $Z_2$ index. The BHZ model has two
    sublattices $A$ and $B$, and we use a truncation where every other
    $A$ sublattice along the $x$ direction is projected out. This
    breaks the invariance of translation by one unit cell along the
    $x$ direction, and effectively doubles the unit cell size, hence
    there are four Wannier spectral lines (instead of two). Black
    dots: Wannier flow of the untruncated state. Purple circles:
    Wannier flow of the truncated state. (a): Non-trivial phase
    (parameters: $\Delta = 0.3, m = 1.1$). (b): Trivial phase
    ($\Delta = 0.3, m = -1.1$). Note different $y$-scales. }
  \label{fig:bhz}
\end{figure}

In Fig.~\ref{fig:bhz}, we compare the Wannier spectral flow
\cite{Yu11} of the ground state at half filling (black dots) with that
of a truncated state (purple circles). The $Z_2$ index can be
identified \cite{Yu11} as the parity of the number of times the
Wannier spectra cross a given value of Wannier center (a constant
$\gamma$ in Fig.~\ref{fig:bhz}) in the half BZ $k_x \in [0, \pi]$. The
truncated state preserves time reversal symmetry because $A_{\ua}$ and
$A_{\da}$ are time reversal partners, hence it still allows for a
$Z_2$ classification. Fig.~\ref{fig:bhz} shows that the $Z_2$ index is
indeed truncation invariant.

\section{Berry phase of truncated $S=1$ AKLT wavefunctions}
The $S=1$ AKLT wavefunction of $N$ spins with a twisted boundary
phase $\kappa$ is
\begin{gather}
  \label{aklt}
  |\Psi(\kappa)\rangle = (a_N\dg b_1\dg e^{i\kappa} - b_N\dg a_1\dg)
  \prod_{i=1}^{N-1} (a_i\dg b_{i+1}\dg - b_i\dg
  a_{i+1}\dg)|\emptyset\rangle\ ,
\end{gather}
where $a$ and $b$ are Schwinger bosons satisfying
\begin{gather}
  S_i^+ = a_i\dg b_i\quad , \quad S_i^z = \frac{1}{2}(a_i\dg a_i -
  b_i\dg b_i)\quad , \quad a_i\dg a_i + b_i\dg b_i \stackrel{!}{=} 2\ ,
\end{gather}
and $|\emptyset\rangle$ is the boson vacuum. On the $i^{th}$ site, one has
\begin{gather}
  (a_i\dg)^2|\emptyset\rangle = \sqrt{2} |\ua\rangle\quad , \quad
  a_i\dg b_i\dg |\emptyset\rangle = |0\rangle\quad , \quad
  (b_i\dg)^2|\emptyset\rangle = \sqrt{2} |\da\rangle\ .
\end{gather}
Inversion $\mathcal{I}$ is defined as
\begin{gather}
  \mathcal{I} a_i \mathcal{I}^{-1} = a_{N+1-i}\quad , \quad
  \mathcal{I} b_i \mathcal{I}^{-1} = b_{N+1-i}\quad , \quad \mathcal{I} i \mathcal{I}^{-1} = i\ .
\end{gather}
This implies that $|\Psi(\kappa)\rangle$ transforms under inversion as
\begin{gather}
  \label{inv-psi}
  \mathcal{I}|\Psi(\kappa)\rangle = (-1)^N e^{i\kappa}
  |\Psi(-\kappa)\rangle\ .
\end{gather}

When fully expanded, Eq.~\ref{aklt} contains $2^N$ monomials by
selecting one of the two terms
$\{a_i\dg b_{i+1}\dg, b_i\dg a_{i+1}\dg\}$ from each bond
\footnote{There are only $2^N - 1$ distinctive terms for even $N$ and
  $2^N - 2$ distinctive terms for odd $N$, depending on whether or not
  the spin configuration $\unexpanded{|0,0,0\cdots, 0\rangle}$ is
  present in the expansion.}. One can verify the spin configurations
corresponding to the monomials in the resulting expansion must satisfy
a string order, wherein the nonzero spins have alternating signs. For
example,
$|1, 0, \cdots, 0, -1\rangle \propto a_1\dg b_2\dg \cdot a_2\dg b_3\dg
\cdots a_{N-1}\dg b_N\dg \cdot b_N\dg a_1\dg |\emptyset\rangle$, and
the (ordered) list of its nonzero spins $\{1,-1\}$ satisfies the
string order.

To gain intuition on the form of projected wavefunctions, consider
first the projection onto a spin configuration $|B_0\rangle$ and its
inversion conjugate $|\bar B_{0}\rangle$, where
\begin{gather}
  |B_0\rangle = |1, 0, 0, \cdots, 0, 0, -1\rangle\quad , \quad
  |\bar B_0\rangle = \mathcal{I}|B_0\rangle = |-1, 0, 0, \cdots, 0, 0,
  1\rangle\ .
\end{gather}
There is only one monomial in the expansion of Eq.~\ref{aklt} that has
non-zero overlap with $|B_0\rangle$,
\begin{gather}
  \langle B_0 | \Psi(\kappa)\rangle = \langle B_0 | a_1\dg b_2\dg \cdot
  a_2\dg b_3\dg \cdots \cdot a_{N-1}\dg b_N\dg \cdot (-b_N\dg a_1\dg)
  | \emptyset\rangle = -2\ , 
\end{gather}
and similarly for $|\bar B_0\rangle$,
\begin{gather}
  \langle \bar B_0 | \Psi(\kappa)\rangle = \langle B_0 | (-b_1\dg
  a_2\dg)\cdot (-b_2\dg a_3\dg) \cdots \cdot (-b_{N-1}\dg a_N\dg)
  \cdot (a_N\dg b_1\dg e^{i\kappa} | \emptyset\rangle = (-1)^{N-1} \times 2e^{i\kappa}\ ,
\end{gather}
which can be alternatively obtained as
$\langle B_0 | \mathcal{I}\Psi(\kappa)\rangle$. The resulting normalized
projected wavefunction is thus
\begin{gather}
  |\wt\Psi(\kappa)\rangle_{B_0} = \frac{1}{\sqrt{2}}\left(|B_0\rangle + (-1)^N e^{i\kappa} |\bar B_0\rangle\right)\ .
\end{gather}
The factor $(-1)^N$ arises due to the inversion conjugacy between
$|B_0\rangle$ and $|\bar B_0\rangle$.

One observes from the above example that the phase in the wavefunction
coefficient of $|B_0\rangle$ depends only on which term in the
boundary link, $a_N\dg b_1\dg e^{i\kappa}$ or $ - b_N\dg a_1\dg$, is
present in the monomial. If the leftmost nonzero spin in a
configuration $|B_0\rangle$ is $1$, then string order demands that
$-b_N\dg a_1\dg$ be present, and the corresponding wavefunction
coefficient is purely real, whereas if it is $-1$,
$a_N\dg b_1\dg e^{i\kappa}$ will be present, and the corresponding
wavefunction coefficient has a phase $e^{i\kappa}$. This observation
is in fact true for any string ordered configuration $|B\rangle$, and
the projected state is
\begin{gather}
  |\wt\Psi(\kappa)\rangle_B = \frac{1}{\sqrt{2}} \left( |B\rangle +
    (-1)^N e^{is\kappa} |\bar B\rangle\right)\ ,
\end{gather}
where $s$ is the leftmost nonzero spin in configuration $B$.

\section{Quench protocol for partial tomography}
In a time-of-flight measurement of a Fermi gas released from an
optical lattice, the momentum distribution is \cite{Bloch08}
\begin{gather}
  \label{n-def}
  n(\vec k) \propto |\wt w(\vec k)|^2 \mathcal{G}(\vec k)\ ,  
\end{gather}
where $\wt w(\vec k)$ is the Fourier transform of Wannier functions
and $\mathcal{G}(\vec k)$ is the Fourier transform of the one-particle
correlation matrix at the time of release,
$\mathcal{G}(\vec k) = \sum_{\vec R, \vec R\pr} e^{i\vec k \cdot (\vec
  R - \vec R\pr)} \langle c\dg_{\vec R}c\ndg_{\vec R \pr}\rangle$.
Here, $c\dg_{\vec R}$ is the creation operator at lattice site
$\vec R$. If there is a sublattice structure, $\vec R$ becomes a
composite label, $\vec R = \vec X + \vec r_a$, where $\vec X$ is the
spatial coordinate associated with the center of a unit cell, and
$\vec r_a$ is the position of the $a^{th}$ sublattice within a unit
cell. Using the Fourier transform
$c_{\vec k, a}\dg = e^{i\vec k \cdot \vec r_a} \left[\sum_{\vec X}
  e^{i\vec k\cdot \vec X}c_{\vec X, a}\dg\right]$, the correlation
matrix becomes
$\mathcal{G}(\vec k) = \sum_{a,b} \langle c_{\vec k, a}\dg c_{\vec k,
  b}\ndg\rangle$.  The correlator is evaluated with the many-body
state $|\Psi\rangle$ of the Fermi gas at the time of release. For a
filled Bloch band,
$|\Psi\rangle = \prod_{\vec q \in BZ} \psi_{\vec q}\dg
|\emptyset\rangle$, where $\psi_{\vec q}\dg$ creates a Bloch band
state of momentum $\vec q$:
$\langle \emptyset | c_{\vec X,a} \psi_{\vec q}\dg |\emptyset\rangle
\equiv e^{i\vec q \cdot (\vec X + \vec r_a)}\psi_a(\vec q)$, and
$\psi_a(\vec q)$ is the Bloch cell function on sublattice
$a$. Knowledge of $\psi_a(\vec q)$ for all $\vec q$ and $a$ would
allow the calculation of the topological index of the single particle
Bloch band
$|\psi(\vec q)\rangle = (\psi_1(\vec q), \psi_2(\vec q), \cdots,
\psi_{N_B}(\vec q))^t$, where $N_B$ is the number of sublattices
within a unit cell. Using Wick's theorem, one then has
\begin{gather}
  \label{gk}
  \mathcal{G}(\vec k) = \sum_{a,b} \langle \emptyset | \psi_{\vec k}\ndg
  c_{\vec k, a}\dg c_{\vec k, b}\ndg \psi_{\vec k}\dg
  |\emptyset\rangle = \left|\sum_a \psi_a(\vec k)\right|^2\ .
\end{gather}
Following Ref.~\onlinecite{Hauke14}, we will ignore the Wannier
envelope $\wt w(\vec k)$ in the momentum distribution, and treat the
correlator $\mathcal{G}(\vec k)$ itself as the momentum
distribution. This is justified because in the quench protocal to be
discussed below, $\wt w(\vec k)$ does not pick up a time dependence,
and since everything of interest will turn out to depend on a ratio,
the $\wt w(\vec k)$ dependence will drop out. Here we also assume that
all atomic basis states originate from the same orbital, e.g., the $s$
orbital. If basis states arise from different orbitals, their Wannier
envelopes will not cancel each other in the way described above
\cite{WeitenbergPrivate}.

From Eq.~\ref{gk}, different wavefunction components (labeled by $a$)
are intermixed in the momentum distribution and thus cannot be
distinguished from each other. The key insight of
Refs.~\onlinecite{Hauke14, Flaschner16a} is that they can be separated
in time domain if the state $|\Psi\rangle$ is subjected to a quench,
for a duration $t_h$ before the ToF measurement, by a flat band
Hamiltonian
$H_{\text{FB}} = \sum_{\vec k, a} \varepsilon_a c_{\vec k, a}\dg c_{\vec k,
  a}\ndg$ between $0 < t < t_h$. $H_{\text{FB}}$ can be achieved by ``turning
off'' electron hopping, and bias different sublattices at different
potentials $\varepsilon_a$. As a consequence, each wavefunction
component will pick up a distinctive dynamical phase at the end of the
quench,
$\psi_a(\vec k, t_h) = \psi_a(\vec k) e^{-i\varepsilon_a t_h}$. The
electrons are then released for a time-of-flight measurement. The
resulting momentum distribution is thus
\begin{gather}
  n(\vec k, t_h) = \left| \sum_a \psi_a(\vec k)
    e^{-i\varepsilon_a t_h} \right|^2\ ,
\end{gather}
note that we have dropped the $\wt w(\vec k)$ dependence as discussed
before, and rescaled to a dimensionless $n(\vec k, t_h)$, see also the
Supplementary Material of Ref.~\onlinecite{Hauke14}.

In general, $n(\vec k, t_h)$ will contain terms oscillating at
frequencies $\omega_{a,b} = |\varepsilon_a - \varepsilon_b|$ due to
the interference between different sublattices. For $N_B$ sublattices,
there are $N_{\omega} = N_B (N_B-1)/2$ such frequencies (assuming no
degeneracy in $\omega$), hence there are $2N_{\omega} + 1$ real
Fourier coefficients $A_{\omega}, B_{\omega}$:
$n(\vec k, t_h) = A_0(\vec k) + \sum_{\omega} [A_{\omega}(\vec k)
\cos(\omega t_h) + B_{\omega}(\vec k) \sin(\omega t_h)]$. For a full
tomography of $|\psi(\vec k)\rangle$, one needs to deduce $2N_B-1$
real-valued unknowns---corresponding to the real and imaginary parts
of the $N_B$ wavefunction components sans the normalization
constraint---from the $2N_{\omega}+1$ experimentally accessible
Fourier coefficients. It is easy to check that
$2N_{\omega} + 1 \ge 2 N_B - 1$, for $N_B \ge 2$, where equality
occurs for $N_B = 2$. That is, we always have enough Fourier
coefficients to fully determine all wavefunction components, hence a
full tomography of a band wavefunction is in principle always
achievable for any number of sublattices. That we have more than
enough Fourier coefficients simply means some of them are not
independent. In practice, however, analytical determination of all
wavefunction components becomes untractable with increasing $N_B$. As
shown in the text, such a full tomography is also unnecessary for
determining the topological index of a wavefunction, for which a
partial tomography of a small subset of wavefunction components would
be sufficient. Below, we discuss a quench protocol for partial
tomography of two wavefunction components. Note that a full tomography
can also be built up from successive partial tomographies.

To perform a partial tomography on, say, the first two sublattices
$a = 1,2$, we set the flat band energy of all other sublattices to a
common level, $\varepsilon_{a>2} = E$, and require that
$\varepsilon_1 \neq \varepsilon_2 \neq E$. The momentum distribution
becomes
\begin{gather}
  n(\vec k, t_h) = \left| \psi_1(\vec k) e^{-i\varepsilon_1 t_h} +
    \psi_2(\vec k) e^{-i\varepsilon_2 t_h} + \left(\sum_{a=3}^{N_B}
      \psi_a(\vec k) \right) e^{-iE t_h} \right|^2\ .
\end{gather}
Introduce the following parametrization,
\begin{gather}
  \psi_1(\vec k) = u(\vec k) \sin \frac{\theta(\vec k)}{2}\quad , \quad \psi_2(\vec k) = -u(\vec k) \cos \frac{\theta(\vec k)}{2} e^{i\varphi(\vec k)}\quad , \quad \sum_{a=3}^{N_B} \psi_a(\vec k) = v(\vec k) e^{i\chi(\vec k)}\ ,
\end{gather}
where $u(\vec k) > 0$, $v(\vec k) \ge 0$, $\theta(\vec k) \in [0,\pi]$, and
$\varphi(\vec k), \chi(\vec k) \in [0, 2\pi]$. Further introduce three frequencies,
\begin{gather}
  \omega_1 = \varepsilon_1 - E\quad , \quad \omega_2 = \varepsilon_2 - E\quad , \quad \omega_3 = \varepsilon_2 - \varepsilon_1\ .
\end{gather}
Then $n(\vec k, t_h)$ has the
following Fourier decomposition (suppressing the $\vec k$ dependence),
\begin{gather}
  n(t_h) = A_0 + \sum_{i=1}^3 \left[A_i \cos(\omega_i t_h) + B_i \sin(\omega_i t_h) \right]\ , \\
  A_0 = u^2 + v^2\ , \\
  A_1 = 2u v \sin \frac{\theta}{2}\cos\chi\quad , \quad B_1 = -2uv\sin \frac{\theta}{2}\sin\chi\ , \\
  A_2 = -2uv\cos \frac{\theta}{2} \cos(\varphi - \chi)\quad , \quad B_2 = -2uv \cos \frac{\theta}{2}\sin(\varphi - \chi)\ ,\\
  A_3 = -u^2 \sin\theta \cos\varphi\quad , \quad B_3 =
  -u^2\sin\theta\sin\varphi\ .
\end{gather}
Note that if $v=0$, $A_{1,2} = B_{1,2} = 0$ and we recover the
$2$-component formalism of Ref.~\onlinecite{Hauke14}. In general,
$v\neq 0$, and the Bloch sphere angles $\theta$ and $\varphi$ can be
determined as
\begin{gather}
  \tan\varphi =  \frac{B_3}{A_3}\quad , \quad \tan \frac{\theta}{2} = \sqrt{\frac{A_1^2 + B_1^2}{A_2^2 + B_2^2}}\ .
\end{gather}
The overall scale $u$ can be obtained as
$u = \sqrt[4]{(A_3^2 + B_3^2)/\sin^2\theta}$, although it does not
enter the evaluation of topological indices such as the Chern number
or the Berry phase. Note that (1) $\varphi$ and $\theta$ only depend
on ratios of the Fourier coefficients, and remain unchanged even when
the Wannier envelope $\wt w(\vec k)$ (cf.~Eq.~\ref{n-def}) is
reinstated, and (2) there are other equivalent expressions for
$\varphi$ and $\theta$ due to the Fourier coefficients not entirely
independent of each other, as discussed before; for example, one can
verify that $\theta$ can be obtained alternatively by
$\tan \frac{\theta(\vec k)}{2} = \left[\frac{B_1(\vec k)}{A_2(\vec k)}
  \sin \varphi(\vec k) - \frac{A_1(\vec k)}{A_2(\vec k)}
  \cos\varphi(\vec k) \right]$.

See Fig.~\ref{fig:hof} for the $\varphi$ and $\theta$ plots resulting
from a 2-sublattice truncation of a Hofstadter band.

\section{Relation with node structure in overlaps of topological wavefunctions}
Recent works \cite{Gu16, Huang16} have shown that if two topological
wavefunctions in the same symmetry class,
$|\Psi_1(\vec \kappa)\rangle$ and $|\Psi_2(\vec \kappa)\rangle$, have
nonzero overlaps in the entire parameter space of $\vec\kappa$, then
they must have the same topological index. Hereafter, we refer to this
as the ``no-node'' theorem, and discuss its relation with the
truncation invariance of topological indices.

We first note that truncation invariance of topological indices is
consistent with the no-node theorem. Consider a topological state
$|\Psi(\vec \kappa)\rangle$ and its truncation
$|\wt \Psi(\vec \kappa)\rangle = P|\Psi(\vec \kappa)/\sqrt{\langle
  \Psi(\vec \kappa) | P | \Psi(\vec \kappa)\rangle}$. In the text we
have shown that $|\wt \Psi\rangle$ and $|\Psi\rangle$ have the same
index as long as
$P|\Psi(\vec \kappa)\rangle \neq 0 \forall \vec\kappa$ and $P$
preserves the protecting symmetry. One can also explicitly verify that
$\langle \wt \Psi | \Psi\rangle$ has no node, because
$\langle \wt \Psi | \Psi\rangle \propto \langle \Psi | P | \Psi\rangle
> 0$ due the nonnegative-definitesess of projection operators (and we
have ruled out $P|\Psi\rangle = 0$). Hence truncation invariance is
consistent with the no-node theorem.

The no-node theorem, however, cannot be used to prove that
$|\Psi\rangle$ and $|\wt \Psi\rangle$ have the same index. This is
because the theorem requires both participating wavefunctions to be
``gapped states''. In Refs.~\onlinecite{Gu16, Huang16}, this condition
is satisfied because both states are explicitly obtained as gapped
ground states of certain physical Hamiltonians. Without first
establishing the ``gapfulness'' of \emph{both} states, the theorem
would not work. Consider for example the BHZ Hamiltonian,
Eq.~\ref{h-bhz}. At $\Delta = 0$, the two spin components are
decoupled, and the lower two bands, $|\psi_{\ua}(\vec k)\rangle$ and
$|\psi_{\da}(\vec k)\rangle$, are degenerate. By construction,
$|\psi_{\ua}\rangle$ and $|\psi_{\da}\rangle$ have opposite Chern
numbers $C = 2s_z = \pm 1$ in the $Z_2$ phase. A generic linear
combination
$|\phi(\vec k)\rangle = \sqrt{f(\vec k)}|\psi_{\ua}(\vec k)\rangle +
\sqrt{1-f(\vec k)} |\psi_{\da}(\vec k)\rangle$, while still an energy
eigenstate, no longer has a quantized Chern number. Now if
$f(\vec k) \neq 0 \forall \vec k$, the overlap of
$|\phi(\vec k)\rangle$ with $|\psi(\vec k)\rangle$ does not vanish
anywhere in the BZ, yet clearly they have different Chern numbers by
construction. This example illustrates the importance of establishing
the ``gapfulness'' before the no-node theorem can be used. In the
investigation of truncation invariance, while we always take a parent
state $|\Psi\rangle$ as a gapped eigenstate of a Hamiltonian, it is
not \emph{a priori} clear whether or not the truncated state
$|\wt \Psi\rangle$ is ``gapped''. Therefore one cannot deduce
truncation invariance from the no-node theorem.

\section{Fractional Chern Insulator}
In the main text, we discussed the implication of Hilbert space
truncation on parton construction and showed that the topological
index computed in the twisted boundary phases of the parent state does
not change after truncation. Here, we perform a direct numerical
calculation to demonstrate the invariance of topological index in
fractional quantum Hall states on a lattice model (also known as
fractional Chern insulator), which host intrinsic topological order in
topologically protected degenerate ground states manifold.

\begin{figure}[t]
  \includegraphics[width=0.35\textwidth]{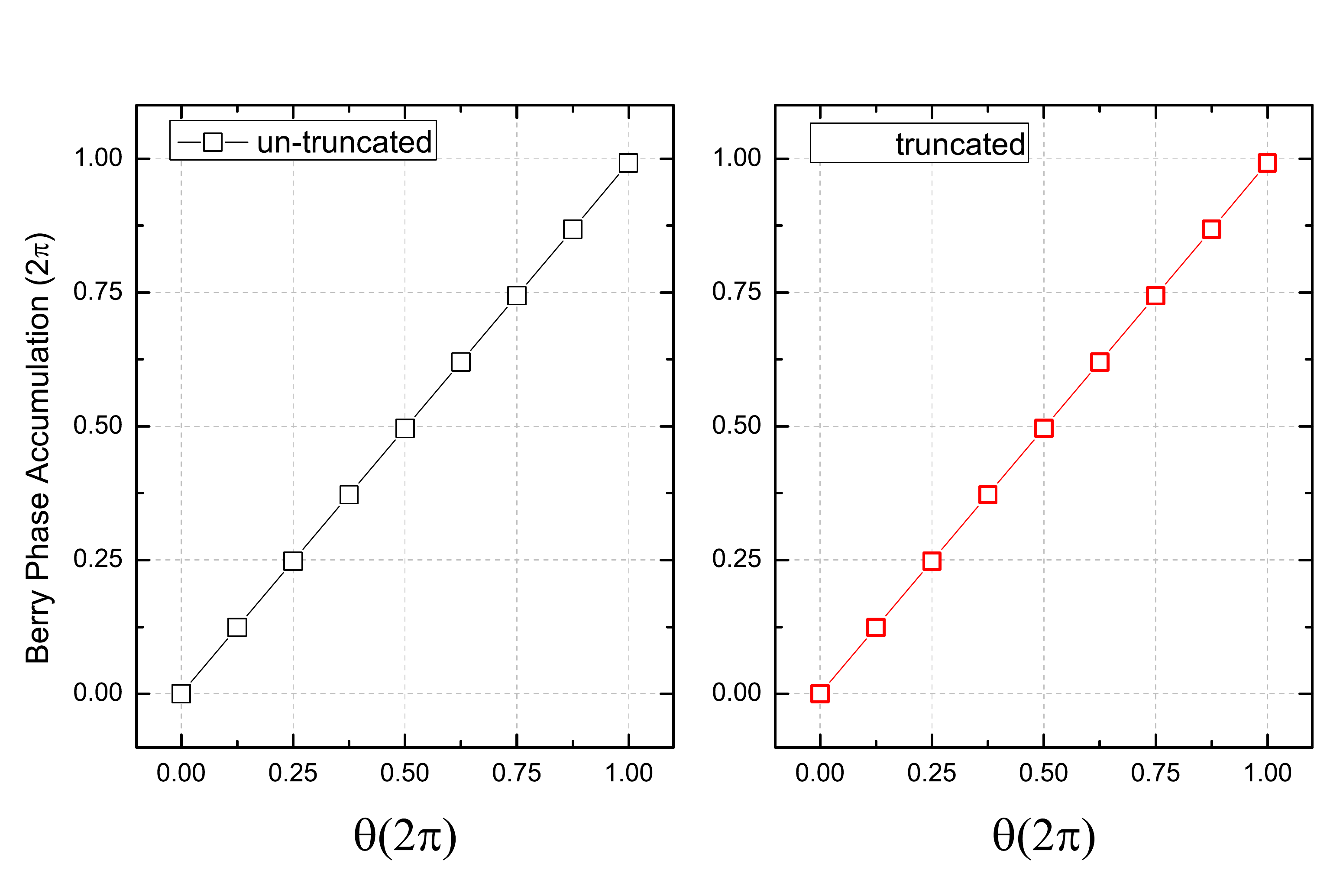}
  \caption{Winding of the accumulated Berry phase as a function of
    $\theta_x$, for (a) the untruncated ground state of $\nu=1$
    non-Abelian Moore-Read state and (b) the corresponding truncated
    ground state. The accumulated Berry phase is defined in
    Eq.~\ref{accu-berry}. Calculation is done with a flat-band model
    on a $2\times 3\times 4$ honeycomb lattice.
    \cite{YFWang12}}\label{fig:fqhe_NA}
\end{figure}

We use a specific topological flat-band lattice model as an example
\cite{YFWang12}, 
where a robust non-Abelian Moore-Read state exists at $\nu=1$.
The Chern number of the many-body ground states can be calculated in
the space of twisted boundary phases $\theta_x$ and $\theta_y$,
\begin{equation}
  C=\frac{1}{2\pi}  \int\limits_0^{2\pi} d\theta_x \int\limits_0^{2\pi} d\theta_y F(\theta_x,\theta_y) \quad , \quad  F(\theta_x,\theta_y)=\Im \left[\langle \frac{\partial \Psi}{\partial \theta_x}|\frac{\partial \Psi}{\partial \theta_y} \rangle - \langle \frac{\partial \Psi}{\partial \theta_y}|\frac{\partial \Psi}{\partial \theta_x} \rangle \right]\ ,
\end{equation}
where $F$ is the Berry curvature. The Chern number is equivalent to
the winding number of the accumulated Berry phase $\gamma(\theta_x)$,
\begin{gather}
  \label{accu-berry}
\gamma(\theta_x) = \int\limits_0^{\theta_x} d\theta_x\pr \int\limits_0^{2\pi} d\theta_y F(\theta_x\pr, \theta_y) \quad , \quad  C = \frac{1}{2\pi} \int d\theta_x \partial_{\theta_x} \gamma(\theta_x)\ .
\end{gather}
For the $\nu=1$ Moore-Read state, there are three quasidegenerate
ground states: a doublet pair in momentum sector $(K_x,K_y)=(0,0)$ and
a singlet in momentum sector $(K_x,K_y)=(0,\pi)$. We truncate to half
of the many-body basis states, and have verified that the
post-truncation Chern number remains invariant regardless of the
truncation basis used. Result from one particular truncation basis is
shown in Fig. \ref{fig:fqhe_NA}, where we plot the winding of the
accumulated Berry phase before (left panel) and after (right panel)
truncation, using the singlet state at $(K_x,K_y)=(0,\pi)$. Before
truncation, the total Berry flux over the whole Brilluin zone is
$2\pi$ within numerical precision, therefore the Chern number is
$C=1$.  The accumulated Berry phase $\gamma(\theta_x)$ of the
truncated state is almost the same as that of the parent state, and
the post-truncation Chern number remains quantized to $C = 1$.

\end{document}